\documentclass[iop,revtex4]{emulateapj} 
\usepackage{pdflscape}
\usepackage{epsfig}
\usepackage{natbib}
\usepackage{graphicx}
\usepackage{amssymb}
\usepackage{amsmath}
\usepackage{multirow}
\usepackage{subfigure}
\shorttitle{}
\shortauthors{Mao et al.}

\begin{document}
\title{Properties of the Magneto-ionic Medium in the Halo of M 51 revealed by Wide-band Polarimetry}

\author{S. A. Mao\altaffilmark{1,2,3,8},
E. Zweibel\altaffilmark{2,4,5},
A. Fletcher\altaffilmark{6},
J. Ott\altaffilmark{1},
F. Tabatabaei\altaffilmark{7}
}

\altaffiltext{1}{National Radio Astronomy Observatory, P.O. Box O, Socorro, NM 87801, USA}
\altaffiltext{2}{Department of Astronomy, The University of Wisconsin, Madison, WI 53706, USA}
\altaffiltext{3}{Max-Planck-Institut f\"{u}r Radioastronomie, Auf dem H\"{u}gel 69, 53121 Bonn, Germany; mao@mpifr-bonn.mpg.de}
\altaffiltext{4}{Department of Physics, The University of Wisconsin, Madison, WI 53706, USA}
\altaffiltext{5}{Center for Magnetic Self-Organization in Laboratory and Astrophysical Plasmas, The University of Wisconsin, Madison, WI 53706, USA}
\altaffiltext{6}{School of Mathematics and Statistics, University of Newcastle, Newcastle upon Tyne, NE1 7RU, UK}
\altaffiltext{7}{Max-Planck-Institut f\"{u}r Astronomie, K\"{o}nigstuhl 17, D-69117 Heidelberg, Germany}
\altaffiltext{8}{S.~A.~Mao is a Jansky Fellow of the National Radio Astronomy Observatory}

\begin{abstract}
We present a study of the magneto-ionic medium in the Whirlpool galaxy (M51) using new wide-band multi-configuration polarization data at  L band (1-2 GHz) obtained at the Karl G. Jansky Very Large Array. By fitting the observed diffuse complex polarization $Q$+$iU$ as a function of wavelength directly to various depolarization models, we find that polarized emission from M51 at 1-2 GHz originates from the top of the synchrotron disk and then experiences Faraday rotation in the near-side thermal halo of the galaxy. Thus, the scale height of the thermal gas must exceed that of the synchrotron emitting gas at L band. The observed Faraday depth distribution at L band is consistent with a halo field that comprises of a plane-parallel bisymmetric component and a vertical component which produces a Faraday rotation of $\sim$ $-$9 rad m$^{-2}$. The derived rotation measure structure functions indicate a characteristic scale of rotation measure fluctuations of less than 560 pc in the disk and approximately 1 kpc in the halo. The outer scale of turbulence of 1 kpc found in the halo of M51 is consistent with superbubbles and the Parker instability being the main energy injection mechanisms in galactic halos.

\end{abstract}


\section{Introduction}
\label{section:introduction}

The origin and evolution of large-scale magnetic fields in galaxies remains an unresolved fundamental question in astrophysics \citep[for a review, see e.g.,][]{kulsrud2008}. Most of our knowledge of galactic-scale magnetic fields come from studies of our own Milky Way. Different probes such as diffuse polarized synchrotron emission \citep[e.g.,][]{carretti2010,wolleben2010}, optical starlight polarization \citep[e.g.,][]{heiles1996}, Zeeman splitting \citep[e.g.][]{green2012} and Faraday rotation towards background extragalactic polarized sources and pulsars \citep[e.g.,][]{vaneck2011,mao2012a,han2006} have been used to infer magnetic field structures in the Galaxy. However, interpreting these observations is difficult due to complicated sight lines through the Galactic plane, our inability to distinguish signatures of large-scale fields from local magnetic features, and pulsar distance uncertainties. Furthermore, extracting the magnetic field structure of the Milky Way from Faraday rotation measures (RMs) of extragalactic sources and pulsars relies heavily on the assumed thermal electron density distribution. Therefore, we turn to external galaxies in order to obtain a clear picture of large-scale magnetic fields in spiral galaxies. 

M51, the Whirlpool galaxy, is a prime candidate for investigating the large-scale magnetic field structures. In fact, it was the first external galaxy detected in linear polarization at radio wavelengths \citep{mathewson1972}. Shorter wavelength observations soon followed \citep{segalovitz1976,beck1987} and led to the tentative suggestion of a bisymmetric magnetic field configuration \citep{tosa1978}. A major step towards understanding the polarized synchrotron emission from M51 was made by \cite{horellou1992}, who derived RM using Very Large Array data at 18 and 20 cm and demonstrated that the  resulting RM magnitude is too small compared to the equipartition estimation of the regular magnetic field strength. The authors concluded that M51 is not transparent in polarization at 18 and 20 cm due to severe Faraday depolarization effects.  

A general picture of M51's large-scale magnetic field emerges from modeling works by \cite{berkhuijsen1997} and \cite{fletcher2011} using data at several sparsely sampled wavelengths (3, 6, 18 and 20 cm). 
Both found that the galaxy hosts distinct large-scale magnetic fields in its disk and halo. Then \cite{fletcher2011} deduced that the disk field is dominated by an axisymmetric mode, whereas the halo field is dominated by a bisymmetric mode. Based on polarization observations of a sample of Spitzer Infrared Nearby Galaxies Survey (SINGS) galaxies at 18 and 22 cm, including M51, \cite{braun2010} proposed that the observed characteristic polarization and Faraday depth modulations as a function of azimuth are produced by an axisymmetric spiral field with a quadrupolar out-of-plane extension in the galaxy's near-side halo.  
M51 was also recently observed at the Giant Metrewave Radio Telescope (GMRT) at 610 MHz \citep{farnes2013} and at the Low-Frequency Array (LOFAR) at 150 MHz \citep{mulcahy2014}, and was found to be completely depolarized within the sensitivity limit. At optical wavelengths, \cite{scarrott1987} showed that linear polarization of M51 forms a spiral pattern from the nucleus out to a radius of 4-5 kpc. However, NIR observations at 1.6$\mu$m conducted using the Mimir instrument on the Perkins telescope show that the galaxy is unpolarized at a level of 0.05\% \citep{pavel2012}. Clearly, much work is needed to converge on a consistent picture of M51's large-scale magnetic field to explain its polarized emission across a wide wavelength range. 

Structure functions of rotation measure can provide information on turbulence in the interstellar medium (ISM). Using RM structure functions of background extragalactic sources behind the Galactic plane, \cite{haverkorn2006,haverkorn2008} derived an energy injection scale of several parsecs (comparable to the typical size of HII regions) inside spiral arms and a scale around 100 pc (comparable to the typical size of supernova remnants) in inter-arm regions. Relatively little is known about the outer scale of turbulence in the diffused ionized gas in external galaxies -- the structure function constructed using RMs of extragalactic polarized sources behind the Large Magellanic Cloud (LMC) led to an estimated scale of 90 pc \citep{gaensler2005}. In M51, \cite{houde2013}, using angular dispersion analysis, derived a turbulent correlation scale
$\sim$ 100 pc in the disk of M51, consistent with the \cite{fletcher2011} result of 50 pc obtained based on the dispersion of RM.

In this paper, we present new L band (1-2 GHz) Karl G. Jansky Very Large Array (VLA) polarization observations of M51 which has the widest frequency coverage compared to previous L band polarization observations of the galaxy. With this broadband dataset in hand, we directly model the measured Stokes $Q$ and $U$ as a function of wavelength to study both small and large-scale structures in the magneto-ionic medium of the galaxy. In Section~\ref{section:observations}, we describe the Jansky VLA observations and data reduction procedures. We present Faraday depths of background extragalactic sources and compute the Milky Way foreground RM in the direction of M51 in Section~\ref{section:extragal_rm}. In Section ~\ref{section:rmsynthesis_diff_emission}, we present the Faraday depth cube obtained from the rotation measure synthesis technique. In Section~\ref{subsection:diffuse_qu_fit}, we present the results of directly fitting to Stokes $Q$ and $U$ to various depolarization models. In Section~\ref{section:nature_faraday_medium}, we interpret results of the $Q$$U$ fitting and discuss the nature of the Faraday rotating medium at L band.  The observed Faraday depth distribution towards M51 is compared to predictions from the \cite{fletcher2011} and the \cite{braun2010} models in Section~\ref{section:large_scale_bfield_model}. In Section~\ref{section:sf_analysis}, the RM structure functions are presented and the outer scales of turbulence in the disk and halo of M51 are extracted.

\section{Observations and Data Reduction}
\label{section:observations}
M51 was observed as part of the science demonstration of the Expanded Very Large Array (EVLA) project. Observations were carried out on 2010 September 6th, 2010 December 5th and 2011 April 10th in D, C and B configurations respectively. Data were taken at L band (971 MHz -- 1995 MHz) in 8 spectral windows, each consisting 64 2-MHz channels. The total time on source was approximately 7 hours. 3C286 was observed at the beginning and the end of each run for absolute flux density, bandpass and polarization angle calibration. Either 3C295 (J1411+5212) or J1313+5458 was used as the complex gain and leakage calibrator and was observed every 30 mins during each run.

Data calibration and reduction were carried out using the Common Astronomy Software Applications\footnotemark[1]\footnotetext[1]{http://casa.nrao.edu} (CASA). Visibilities affected by radio frequency interference (RFI) were flagged using the automated flagging algorithm RFLAG in CASA. The measurement sets were carefully inspected for further RFI excisions. After flagging, approximately 400 MHz of the total 1 GHz bandwidth at L band remain usable. Raw data were then binned into 8 MHz channels before calibration. Flux densities were scaled to the \cite{perley2013a} 3C286 model. The measured polarization angles were calibrated to 3C286, whose polarization angle is +33$^\circ$ across L band  \citep{perley2013b}. Time dependent antennae gains were solved on a per-spectral-window basis, while polarization leakage and absolute polarization angle calibration were solved on a per-channel basis. Following the leakage calibration, the residual instrumental polarization level is below 0.25\%. We have subsequently masked all pixels with fractional polarization below 0.25\%. We have verified, using Faraday rotation synthesis that 3C286 has a Faraday depth consistent with zero ($-$0.02 $\pm$ 0.03 rad m$^{-2}$).

\subsection{Imaging Total Intensity}

After separately calibrating the measurement sets of different array configurations, we have constructed a Stokes $I$ natural-weighted image across L band. We have utilized the multi-scale, multi-frequency deconvolution as implemented in CASA \citep{rau2011} which models the sky brightness as a linear combination of both spectral and spatial basis functions. Two Taylor terms were used to model the spectral dependence. In order to model the extended emission in the field of view, we used angular scales ranging from 0" (which corresponds to point sources) to 5.3' (comparable to the extent of the galaxy). We imaged a 30'$\times$30' field around the phase center. 

\begin{figure*}[h]
\centering
\includegraphics[width=0.9\textwidth]{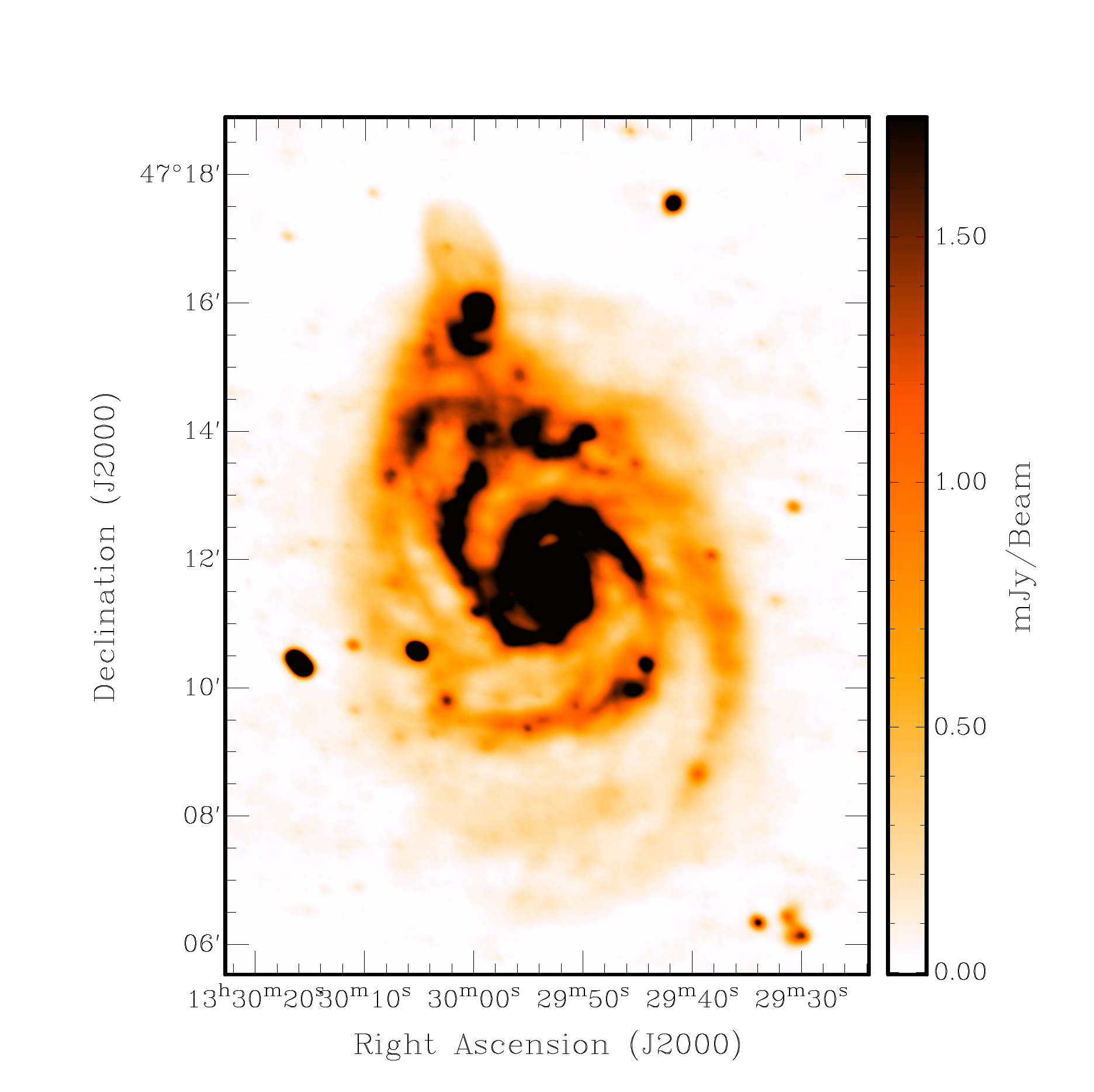}
\caption{Stokes $I$ multi-frequency synthesis map of M51 at 1.478 GHz using the new wide-band multi-configuration VLA data.}
\label{fig:m51_stokesi_map}
\end{figure*}

\begin{figure*}
\centering
\includegraphics[width=0.9\textwidth]{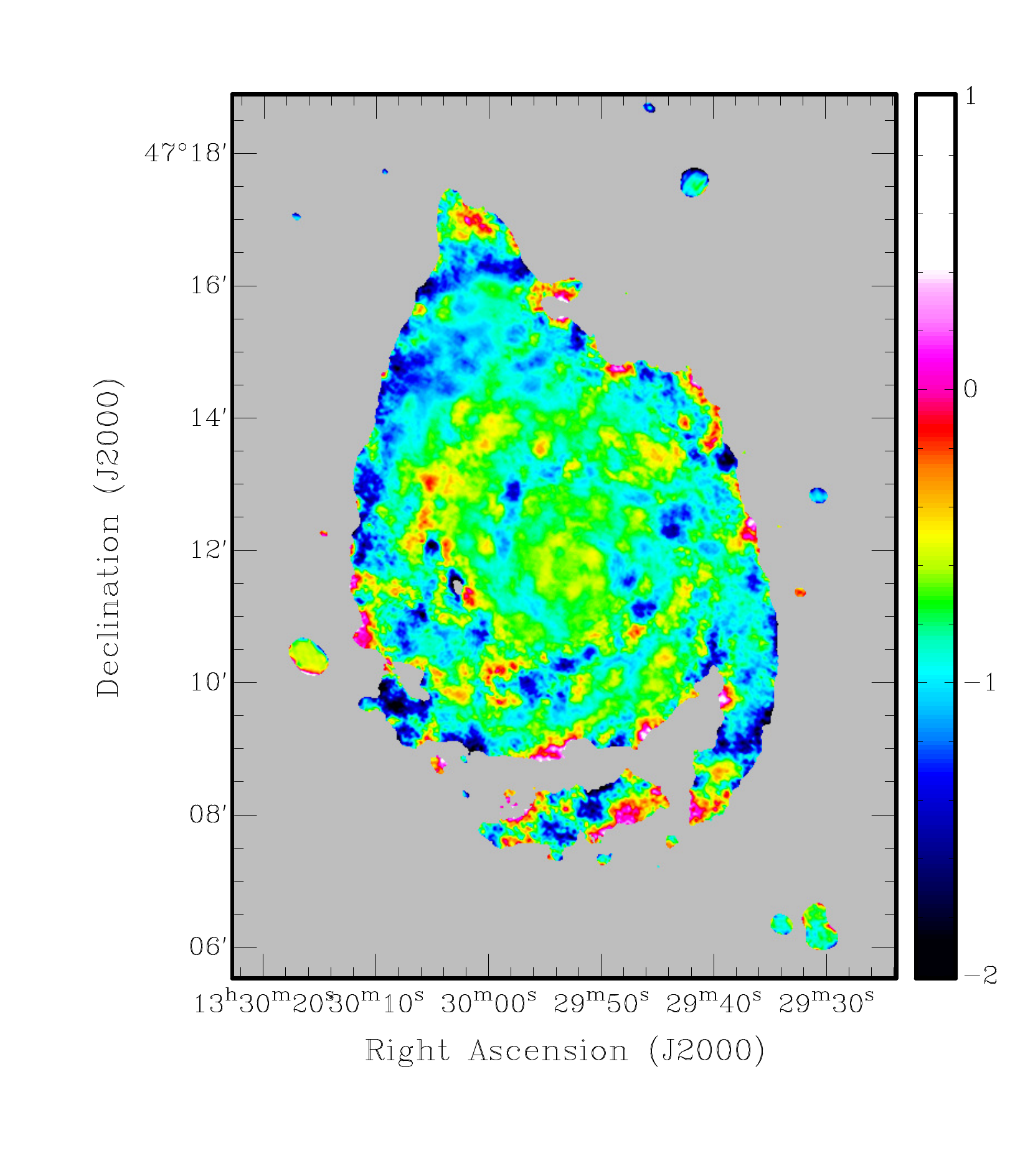}
\caption{Spectral index map of M51 at L band using the new wide-band multi-configuration VLA data. Pixels with signal-to-noise less than 3 have been masked.}
\label{fig:m51_spin_map}
\end{figure*}

The resulting total intensity map, after primary beam correction, is shown in Figure~\ref{fig:m51_stokesi_map}, with a noise level of 23 $\mu$Jy beam$^{-1}$ at a resolution of 10.9''$\times$8.8''. A spectral index\footnotemark[2]\footnotetext[2]{We define spectral index $\alpha$ as I = I$_0$$\nu$$^\alpha$.} map across L band is also produced and is shown in Figure~\ref{fig:m51_spin_map}, with a typical uncertainty of 0.2. The total integrated flux density of M51 is measured to be 1.4 Jy at 1.478 GHz, which is in good agreement with previous single-dish \citep{lequeux1971} and interferometric observations \citep{segalovitz1977,dumas2011}. Therefore, we conclude that no large-scale flux is missing from our new Jansky VLA multi-configuration data.

\subsection{Imaging Stokes $Q$ and $U$}
To facilitate a Faraday rotation study, images of Stokes $Q$ and $U$ are made at each of the 45 8-MHz channels using multi-scale CLEAN, with angular scales spanning from ``zero" to 2.7'. Natural weighting scheme is used in order to maximize the sensitivity to weak and extended polarized emission. These channel maps are first made at their original resolutions and then corrected for the response of the primary beam. Subsequently, these images have all been smoothed to a common resolution of 10.9"$\times$8.8" to match the resolution at the lowest frequency. The typical noise in a Stokes $Q$ or $U$ channel map is about 56 $\mu$Jy beam$^{-1}$, close to the theoretical noise level of about 51 $\mu$Jy beam$^{-1}$.

\subsection{Computing Faraday Depths} 
When linearly polarized light travels through a magnetized medium, its plane of polarization rotates due to the Faraday rotation effect. In the case where background polarized emission experiences pure Faraday rotation in a foreground medium, the angle of rotation in radians is given by 
\begin{equation}
\Delta\psi= {\rm RM}  \lambda^{2}
\end{equation}
where $\lambda$ is the wavelength of the radiation measured in meters and RM is the rotation measure. In the more general case of mixed emitting and rotating medium along the line of sight, or of multiple RM components within the telescope beam, Faraday depth $\phi$ is defined as: 
\begin{equation}
\phi =0.812 \int ^{observer}_{source} {n_{e}(l){B_{\parallel} (l)}} dl~~~\rm{rad~m^{-2}}
\label{vlaeq:rmdef}
\end{equation}
In the above equation, $n_e(l)$ (in cm$^{-3}$) is the thermal electron density, $B_{\parallel}(l)$ (in $\mu$G) is the line of sight magnetic field strength and d$\it{l}$ (in pc) is a line element along the line of sight. 
RM is equivalent to the Faraday depth $\phi$ in the simple case of pure rotation caused by a single component. We note that the electron column density is more fundamental than the volume density because a path length assumption is needed in order to convert column density into volume density.

\begin{figure*}
\centering
\includegraphics[width=0.9\textwidth]{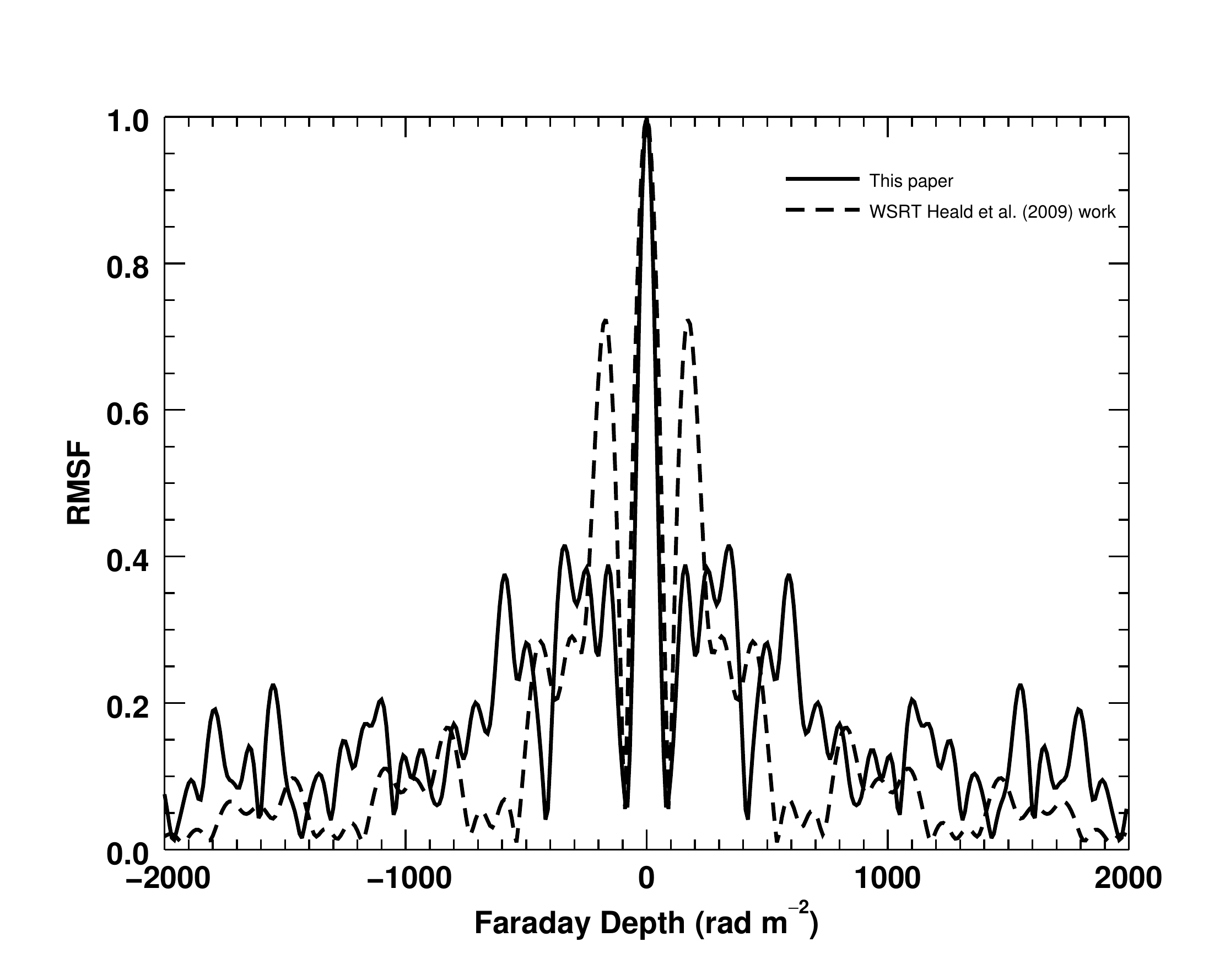}
\caption{Comparison between the rotation measure spread function of our data (solid line) and that of \cite{heald2009} (dashed line). The more continuous frequency coverage of our Jansky VLA data not only provides an improved resolution in Faraday depth space, it also significantly suppresses the amplitude of the first side-lobe from $\sim$ 78\% down to $<$ 35\%.}
\label{fig:compare_rmsf}
\end{figure*}

We have computed the Faraday depth of M51 at L band using the RM synthesis technique \citep{brentjens2005}, followed by the deconvolution of RM spectra using the RMCLEAN algorithm \citep{heald2009}. We stop cleaning when the peak of the spectrum falls below 4 times the noise level. Given the frequency setup of our observations, the rotation measure spread function (RMSF) has a full width at half-maximum (FWHM) of 90 rad m$^{-2}$, a great improvement over the restoring beam of 144 rad m$^{-2}$ of the Westerbork  (WSRT) SINGS observation of M51 \citep{heald2009,braun2010}. Moreover, the much improved frequency coverage of our new VLA observations have significantly suppressed the side-lobe level in the RMSF: the first side-lobe is at a level of approximately 35\% of the main peak (compared to 78\% for the WSRT work). The higher resolution and the much lower side-lobe levels of our Faraday depth spectra enable us to better interpret structures in Faraday depth ($\phi$) space. The deconvolution algorithm will also perform better because high side-lobes are less likely to be mistaken as real signal during RMCLEAN. For illustration, we have plotted the RMSF of our VLA observations and that of the WSRT observations \citep{heald2009} in Figure~\ref{fig:compare_rmsf}. The channel width (8 MHz) limits our sensitivity to Faraday depths $<$ 7400 rad m$^{-2}$. The highest observing frequency (1.835 GHz) sets our sensitivity to extended structure in $\phi$ space: the sensitivity drops by 50\% for structures with $\phi$ extents greater than 118 rad m$^{-2}$.

\section{Results}
\label{section:extragal_rm}

\subsection{Faraday Depths of Polarized Extragalactic Background Sources}

To extract reliable polarization information from extragalactic sources across the field and to avoid confusion with diffuse emission associated with M51, we have re-imaged the data in Stokes $I$, $Q$ and $U$ using only baselines longer than 350 m. We imaged a field of size 40'$\times$40' around the phase center, and thus ``w-projection" was required to correct for non-coplanar effects. 
The resulting images are only sensitive to emission on scales smaller than 2'. The resulting Stokes $I$ map has a noise level of 25 $\mu$Jy beam$^{-1}$, whereas the Stokes Q and U 8-MHz channel maps have an rms noise of about 45 $\mu$Jy beam$^{-1}$ with a beam size of 13.2"$\times$8.7". Coordinates of all polarized sources in the field of view and their distance from the pointing center are listed in Table~\ref{table:egs_catalog}. There are 5 polarized sources within 20' from the pointing center: one source with three components, two doubles and two unresolved sources. We form $Q/I$ and $U/I$ cubes using the total intensity and spectral index map at the reference frequency of 1.478 GHz. 

Although RM Synthesis is an excellent tool to visualize polarized emission at various Faraday depths, there is considerable ambiguity in extracting properties of the underlying Faraday structure using this approach, especially when there is complexity in the spectrum \citep{farnsworth2011,osullivan2012}. The latest RM benchmark test for different Faraday decomposition methods \citep{sun2014b} demonstrates that directly fitting for Stokes $Q$, $U$ as a function of $\lambda^2$ is most successful in recovering the correct components in $\phi$ space. In addition, modeling the depolarization trend can reveal properties of the magneto-ionic medium that are otherwise hard to obtain, such as the relative distribution of synchrotron emitting and Faraday rotating material as well as turbulent properties in the medium.

We first estimate the Faraday depth of each source (or component) at the pixel with the highest signal-to-noise detection in polarization using RM synthesis \citep{brentjens2005}. The result is subsequently used as the initial parameter guess for a maximum likelihood fit to $Q$/$I$, $U$/$I$ as a function of $\lambda^2$ to several models of the synchrotron emitting and Faraday rotating region. We consider the following models (i) a uniform external Faraday rotating screen\footnotemark[3]\footnotetext[3]{The term screen is used throughout this paper to represent a region which consists only of thermal electrons with no relativistic electrons.}; (ii) an inhomogeneous external Faraday screen; (iii) an inhomogeneous Faraday screen with a partial covering fraction; (iv) a Burn slab (consists of well-mixed thermal and synchrotron emitting gas) with a regular magnetic field; (v) a slab with both regular and random fields and (vi) two spatially unresolved Faraday depth components. Expressions corresponding to these models are listed in Appendix~\ref{appendix:A}. In addition to maximum likelihood fits, least-square fits were performed as well. We confirm that the two methods converge to the same solutions. We select a model that provides a satisfactory fit without introducing too many free parameters. More complicated models are accepted only when they significantly improve the reduced $\chi^2$ of the fit (at least at 5$\sigma$ level) based on the F-test. In addition, we utilize the Bayesian information criterion (BIC) defined in \cite{osullivan2012}\footnotemark[4]\footnotetext[4]{We note that the value N in their Equation (14) should be twice the total number of frequency channels.} to select the best-fit model. The BIC penalizes models with a large number of free parameters and at the same time it allows for an easy comparison of non-nested models. We have chosen a criterion for the BIC such that only when BIC$_{\rm model 1}$ $-$ BIC$_{\rm model 2}$ $>$ 30 do we favor the more complex model 2. 

\begin{figure*}
\centering
\includegraphics[width=0.9\textwidth]{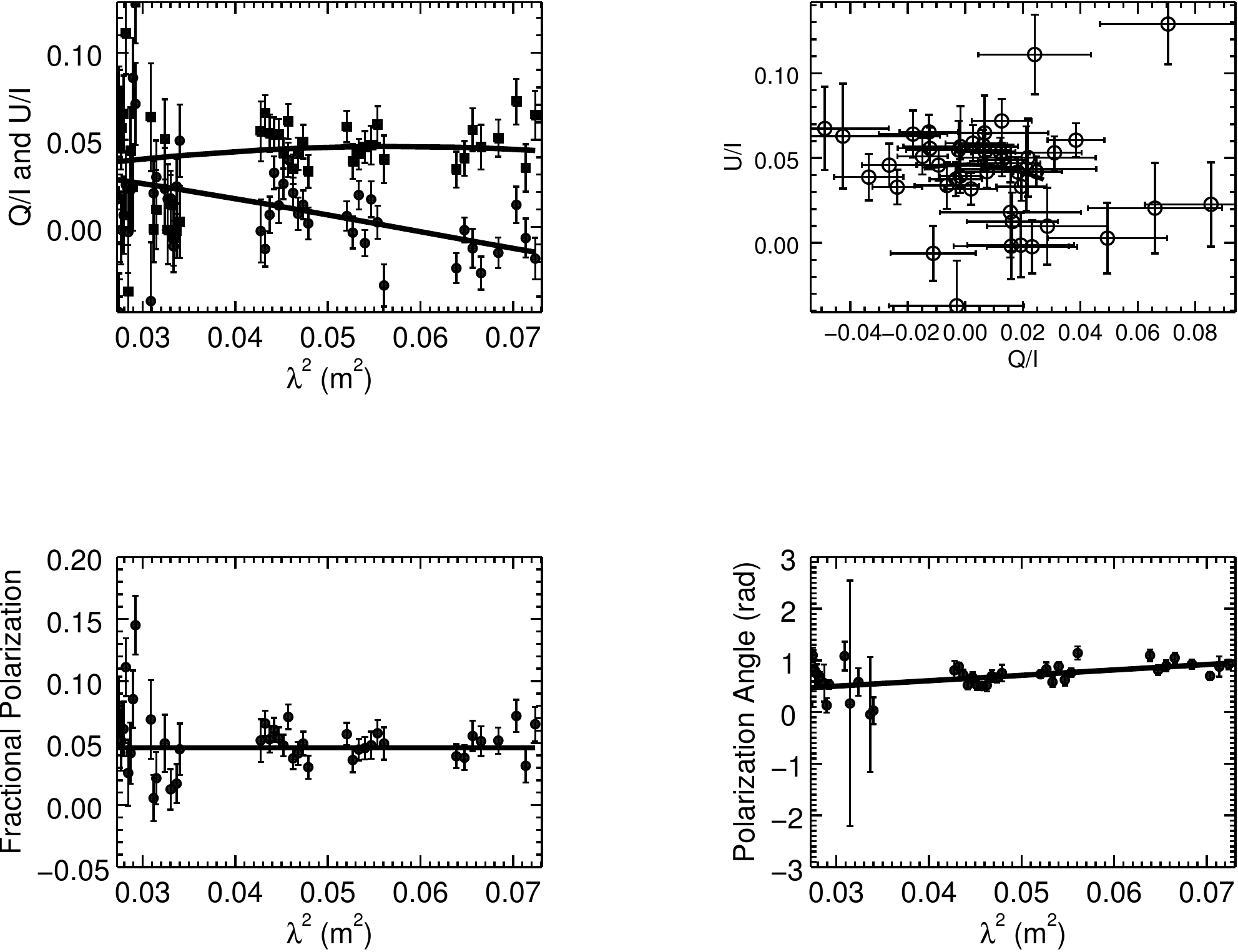}
\caption{Polarized emission of source J1330+4703b and the corresponding best fit (solid line) to the uniform external Faraday screen model. We show the fractional $Q$ (filled squares) and $U$ (filled circles) versus $\lambda^2$ in the top left plot.  The top right plot shows the $Q/I$-$U/I$ track. The bottom left plot shows the fractional polarization versus $\lambda^2$ trend. The bottom right plot shows how polarization angles vary across L band.}
\label{fig:pt_source_1}
\end{figure*}

\begin{figure*}
\centering
\includegraphics[width=0.9\textwidth]{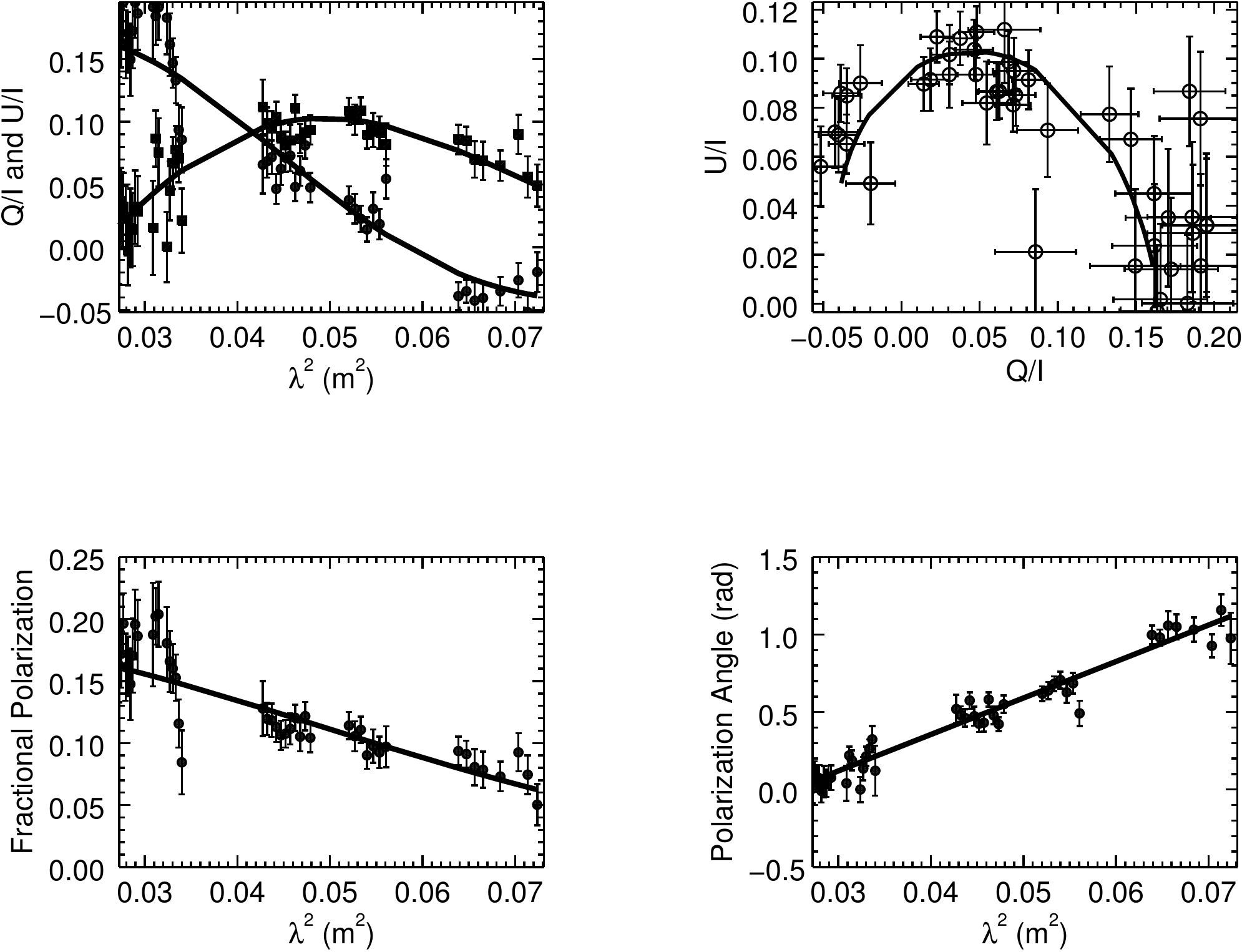}
\caption{Polarized emission of source J1329+4717 and the corresponding best fit (solid line) to the inhomogeneous external Faraday screen model. This figure has the same layout as in Figure~\ref{fig:pt_source_1}.}
\label{fig:pt_source_2}
\end{figure*}

\begin{figure*}
\centering
\includegraphics[width=0.9\textwidth]{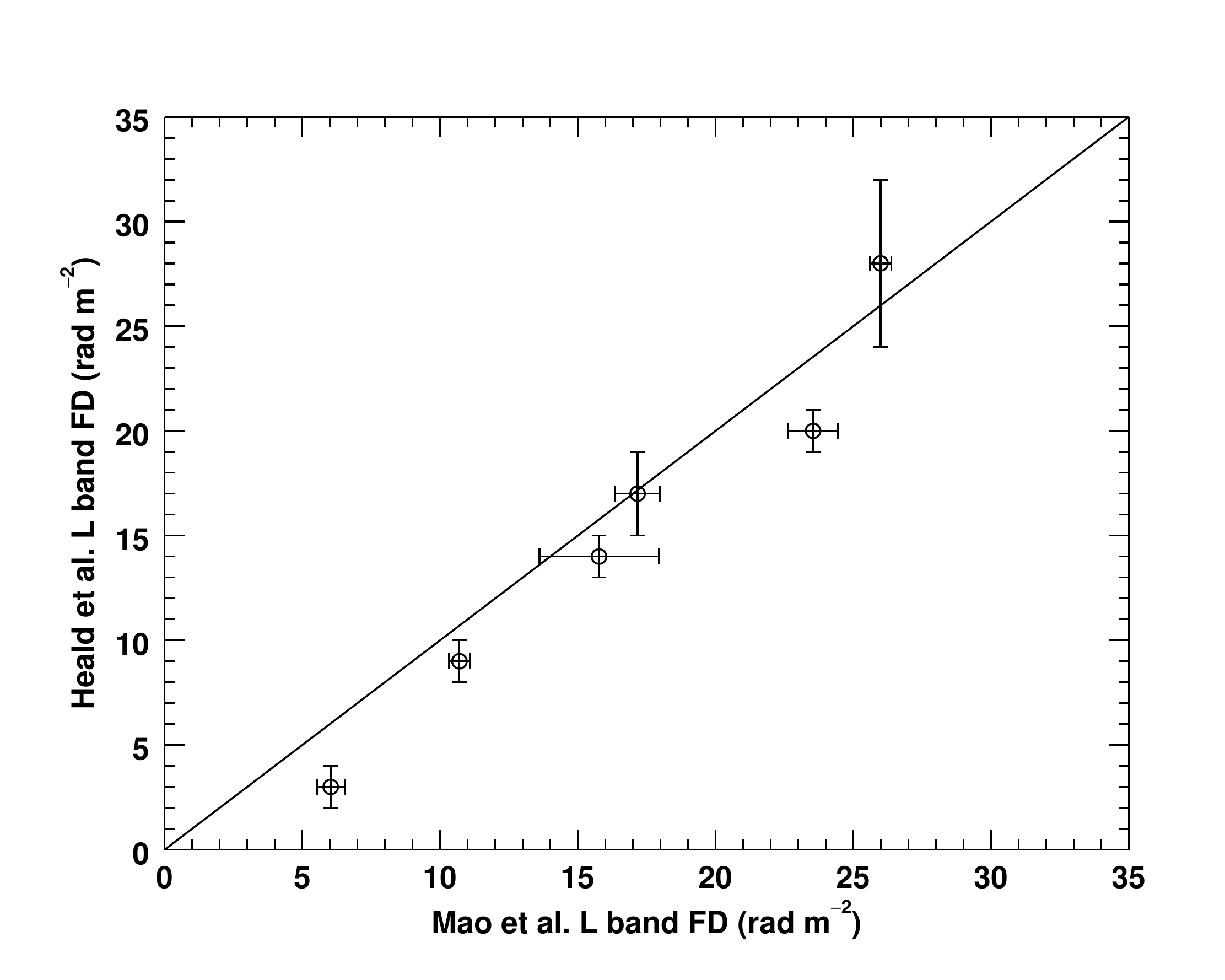}
\caption{Comparison of Faraday depths of extragalactic background sources reported in \cite{heald2009} against the Faraday depths derived from our Jansky VLA L band data. The solid line of slope 1 indicates where \cite{heald2009} and our VLA Faraday depths agree with each other.}
\label{fig:heald_vs_mao}
\end{figure*}

We list the fitted parameters, their uncertainties, $\chi_r^2$ and the BIC for each source (or component) and for each model in Table~\ref{table:pointsource_RM_fit}. The best fit models are indicated using boldface font in the Model column. We show the polarization data and the corresponding best fit models to source J1330+4703b and J1329+4717 in Figure~\ref{fig:pt_source_1} and~\ref{fig:pt_source_2} respectively. Since off-axis frequency-dependent instrumental polarization response was not corrected for in our data set, we choose to only report Faraday depths of sources within 20' from the pointing center. In Figure~\ref{fig:heald_vs_mao}, we plot our derived L-band Faraday depths (FD) against those derived by \cite{heald2009} for extragalactic polarized sources that overlap in both studies. Approximately 60\% of sources in both samples have consistent FDs within their measurement errors. This suggests a good agreement between the two L band datasets.

We list the fractional polarizations and the Faraday depths of sources that are present in both our VLA data set and the \cite{farnes2013} GMRT observations in Table~\ref{table:overlapped_sample}. For these sources, we predict the fractional polarization at 610 MHz using our best-fit models and compare them to the observed values\footnotemark[5]\footnotetext[5]{The percent polarization listed in their Table 1 is an upper limit because off-axis Stokes $I$, $Q$ and $U$ effects were not corrected for.}. For 3 out of the 6 extragalactic sources, predictions of the fractional polarization at 610 MHz from our direct $Q$ $U$ fits are consistent with the observed values. However, we overpredict the fractional polarization for the remaining 3 sources. This could be due to an additional unresolved and weakly-polarized steep-spectrum component that dominates at low frequencies, resulting in the much lower total fractional polarization at 610 MHz. We suggest that a rigorous comparison should be performed only after all off-axis instrumental polarization effects have been calibrated out in both our Jansky VLA and the GMRT \cite{farnes2013} data sets.  Using the 6 extragalactic sources with both 610 MHz and L-band polarization information, we find an average depolarization ratio of 0.4 between 610 MHz and 1.4 GHz. We note that \cite{giessubel2013} found an average depolarization of extragalactic sources around M31 between 325 MHz and 1.4 GHz of $\sim$ 0.14. Since wavelength-dependent depolarization effects are less severe at 610 MHz than at 325 MHz, it is reasonable to obtain a higher depolarization ratio ($>$ 0.14) at 610 MHz.

\subsection{Milky Way Foreground RM in the Direction of M51} 

Previous estimates of the Milky Way foreground rotation measure in the direction of M51 have yielded very different results. 
\cite{horellou1992} estimated a foreground RM of $-$5$\pm$12 rad m$^{-2}$ by averaging RMs of 9 extragalactic sources in the \cite{sn1981} catalog within 20$^\circ$ of M51. In the WSRT M51 work, \cite{heald2009} obtained a foreground RM value of $+$12$\pm$2 rad m$^{-2}$ using the RMs of 5 extragalactic sources detected in the field of M51. 

To determine the Milky Way foreground RM, we only use statistics of polarized extragalactic sources within 20' of M51 detected in our VLA observations to minimize effects from uncorrected off-axis instrumental polarization. 
Moreover, this criterion limits the variance in RM introduced by fluctuations in thermal electron density and magnetic fields in the Milky Way foreground. At the Galactic latitude of M51 ($b$$=$ $+$68$^\circ$), RM variance in the Milky Way remains strong, up to 9 rad m$^{-2}$ \citep{schnitzeler2010}. Since some extragalactic sources are located behind neutral hydrogen tidal features with substantial column densities that could produce enhancements in Faraday rotation with even a low ionization fraction, we further restrict our foreground sight lines to have HI column densities $<$ 10$^{20}$ cm$^{-2}$ in the \cite{rots1990} HI map. The median Faraday depth of the 6 sources (denoted with * in Table~\ref{table:egs_catalog}) meeting these criteria  is $+$13 rad m$^{-2}$ with a standard error of 1 rad m$^{-2}$, which we adopt as the constant Milky Way foreground RM in the direction of M51. Our Milky Way foreground estimate is consistent with the \cite{heald2009} estimation, but it is very different from the \cite{horellou1992} estimation due to the large circle (radius 20$^\circ$) within which RMs of extragalactic sources have been averaged in their study.

\subsection{RM Synthesis Results of the Diffuse Polarized Emission from M51}
\label{section:rmsynthesis_diff_emission}

\begin{figure*}
\centering
\includegraphics[width=0.8\textwidth,clip=true]{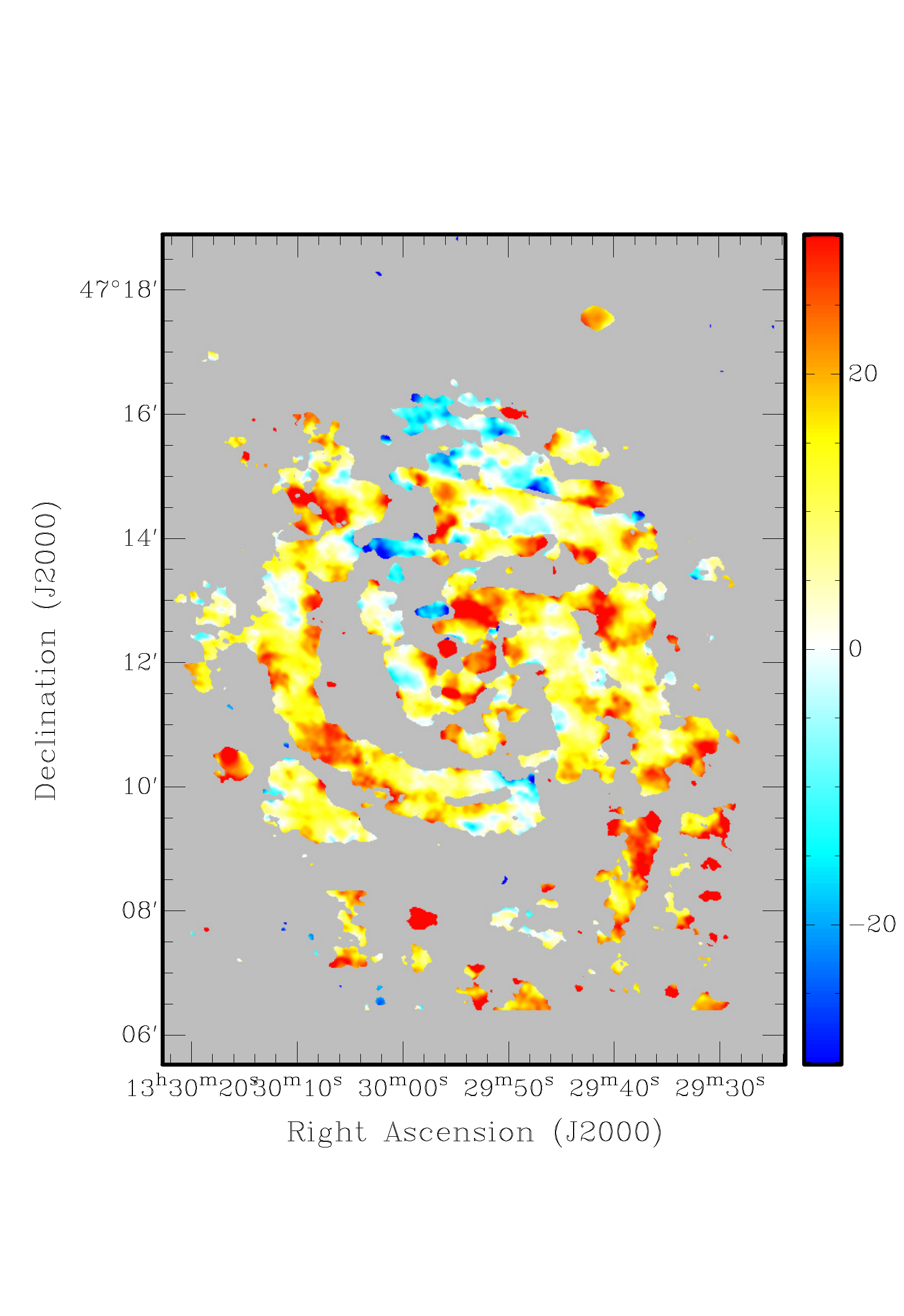}
\caption{Faraday depth distribution of M51 at L band derived from RM synthesis. The color scale is in the unit of rad m$^{-2}$.} 
\label{fig:rm_peak}
\end{figure*}

\begin{figure*}
\centering
\includegraphics[width=0.8\textwidth,clip=true,angle=-90]{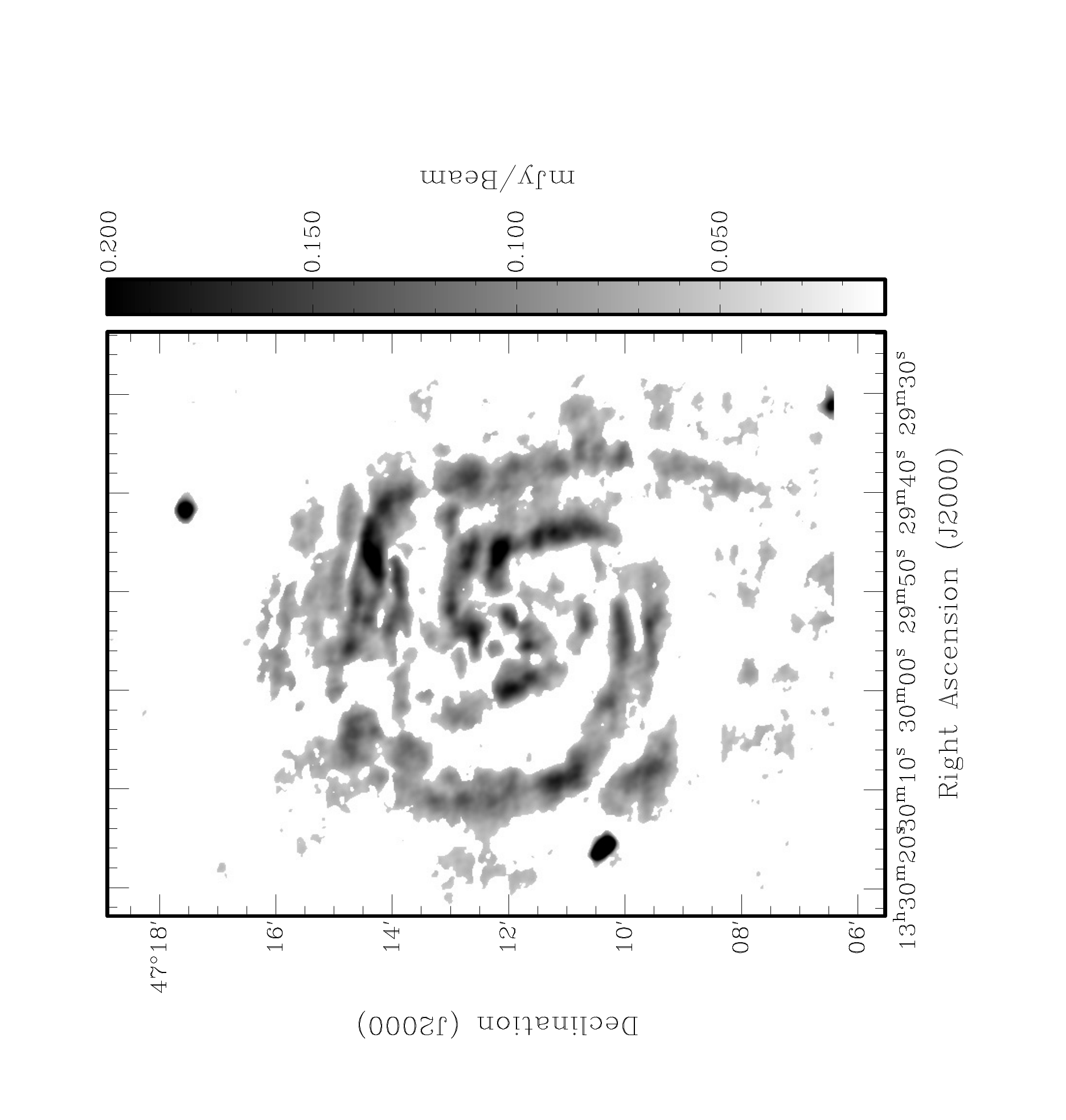}
\caption{The de-biased polarized intensity of M51 at the peak of the Faraday depth spectrum.}
\label{fig:pi_peak}
\end{figure*}

\begin{figure*}
\centering
\includegraphics[width=1.0\textwidth]{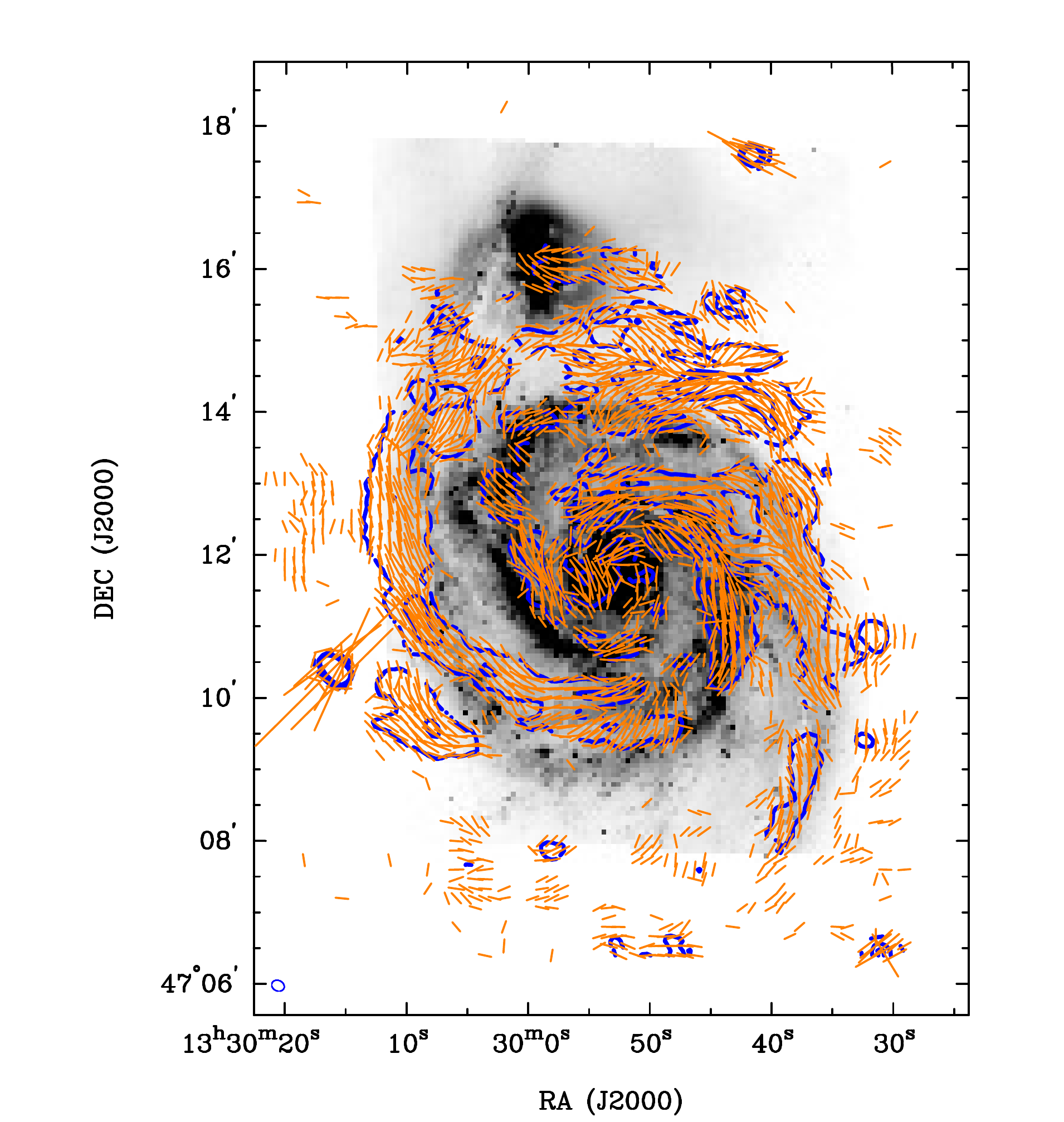}
\caption{Faraday rotation-corrected intrinsic magnetic field orientations (orange line segments) and the polarized intensity contour (blue) at 65 $\mu$Jy beam$^{-1}$ overlaid on a B band Hubble Space Telescope image of M51 \citep{mutchler2005}.}
\label{fig:pa_on_hst}
\end{figure*}

We determine the Faraday depth of diffuse polarized emission from M51 using the RM synthesis technique \citep{brentjens2005}. Along each sight line, we extract the peak polarized intensity and the corresponding Faraday depth from the Faraday depth spectrum. The resulting polarized intensity and Faraday depth distributions across the galaxy are displayed in Figures~\ref{fig:rm_peak} and~\ref{fig:pi_peak} respectively. Only pixels with signal-to-noise detection in polarization greater than 7 are displayed\footnotemark[6]\footnotetext[6]{This is equivalent to 6$\sigma$ in standard Gaussian statistics \citep{hales2012}.}. Assuming a single Faraday depth component along the line of sight, we plot the magnetic field position angles (after correcting for Faraday rotation) on the Hubble Space Telescope B band image \citep{mutchler2005} in Figure~\ref{fig:pa_on_hst}. There appears to be polarized emission which coincides with the companion galaxy NGC 5195 and extends westward. However, the relative distance of M51 and NGC 5195 is required to determine if this polarization is truly intrinsic to NGC 5195. Polarized synchrotron emitting arms and optical arms form a complicated pattern across the galaxy. The relative locations of the magnetic and material arms were investigated in details by \cite{patrikeev2006}. In this paper, we concentrate on the structure of the synchrotron emitting and Faraday rotating material along the line of sight. 
The integrated polarized flux of M51 is measured to be  82 mJy which is in excellent agreement with the value 81$\pm$ 5 mJy reported by \cite{heald2009}. 
We do not find evidence for extended Faraday structures or multiple Faraday depth components in the spectra: there exists no distinct multiple peaks at high significant levels in the spectra nor are there signs of broadening of the main peak.

\subsection{Direct Fit to Q($\lambda^2$) and U($\lambda^2$)}
\label{subsection:diffuse_qu_fit}
We have performed a direct fit to Stokes $Q$ and $U$ as a function of $\lambda^2$ on a pixel-by-pixel basis across M51 using the \cite{osullivan2012} maximum likelihood approach. We note that this is the first time this approach has been applied to wide-band data of extended polarized emission from galaxies: previous works involve fitting $Q(\lambda)$ and $U(\lambda)$ of unresolved extragalactic sources. In order to decouple depolarization from simple spectral index effects, care has to be taken to estimate the synchrotron fractional polarization. Unlike extragalactic radio sources whose emission at L band is mostly non-thermal so that one can simply divide out the observed $Q$ and $U$ by $I$ to correct for spectral index effects, the situation for radio emission from a galaxy is complicated by the contamination of non-negligible free-free emission at L band. Based on radio data spanning frequencies from 0.4 to 22.8 GHz, \cite{klein1984} found the total thermal fraction of M51 at 1.4 GHz to be 5\% with an integrated synchrotron spectral index of $-$0.98 \citep{klein1984}. From the radio spectral index map derived from our VLA data (Figure~\ref{fig:m51_spin_map}) and that presented in \cite{fletcher2011}, we see that the radio spectral index can be as steep as $-1.1$ in parts of M51. Moreover, depending on how far above/below the mid-plane the emission originates, the synchrotron spectral index could be even steeper because of cosmic ray energy losses \cite[e.g.][]{heesen2009}, unless there are localized outflows. Since it is believed that the observed polarized emission at L band originates from the top of the synchrotron emitting disk in the galaxy's near side \citep{fletcher2011,braun2010}, we adopt a rather steep synchrotron spectral index of $-$1.1 for all sight lines. We acknowledge that the synchrotron spectral index likely varies across the face of the galaxy and so for sight lines that have a shallower spectral index, the amount of depolarization would be overestimated. For this reason, we repeated the fitting procedure assuming a synchrotron spectral index of $-$0.5 and the conclusion on the general properties of the Faraday rotating medium in M51 remains unchanged.  

We assume the Milky Way foreground Faraday rotation occurs in a screen in between us and M51, and then consider the following models for M51's magneto-ionic medium: (i) uniform background polarized emission propagates through a uniform external Faraday rotating screen; (ii) uniform background polarized emission propagates through an inhomogeneous external Faraday rotating screen ; (iii) a Burn slab that consists of a regular field and well mixed thermal and cosmic ray electrons; (iv) a slab with both regular and random fields and (v) two spatially unresolved Faraday components. Equations that correspond to these models are listed in Appendix~A. We select the best fit model by comparing the BIC for the fits to all 5 models. We select the least complex model that minimizes the BIC and satisfies BIC$_{\rm model1}$$-$BIC$_{\rm model2}$ $>$ 30. We note that this BIC criteria roughly corresponds to favoring model 2 over model 1 at $\ge$ 6$\sigma$ level according to the F-test. We allowed the Milky Way foreground RM in these models to be a free parameter. Both Model (iii) and (iv) produce best-fit Milky Way foreground RM values in the range of $\sim$ $-$200 rad m$^{-2}$ to $\sim$ +140 rad m$^{-2}$ with large pixel-to-pixel variations. Since model (iii) and (iv) both yield Milky way foreground RM values inconsistent with our estimation of $+$13 $\pm$1 rad m$^{-2}$, we do not consider these two models any further. 

\begin{figure*}
\centering
\includegraphics[width=0.8\textwidth,clip=true,angle=-90]{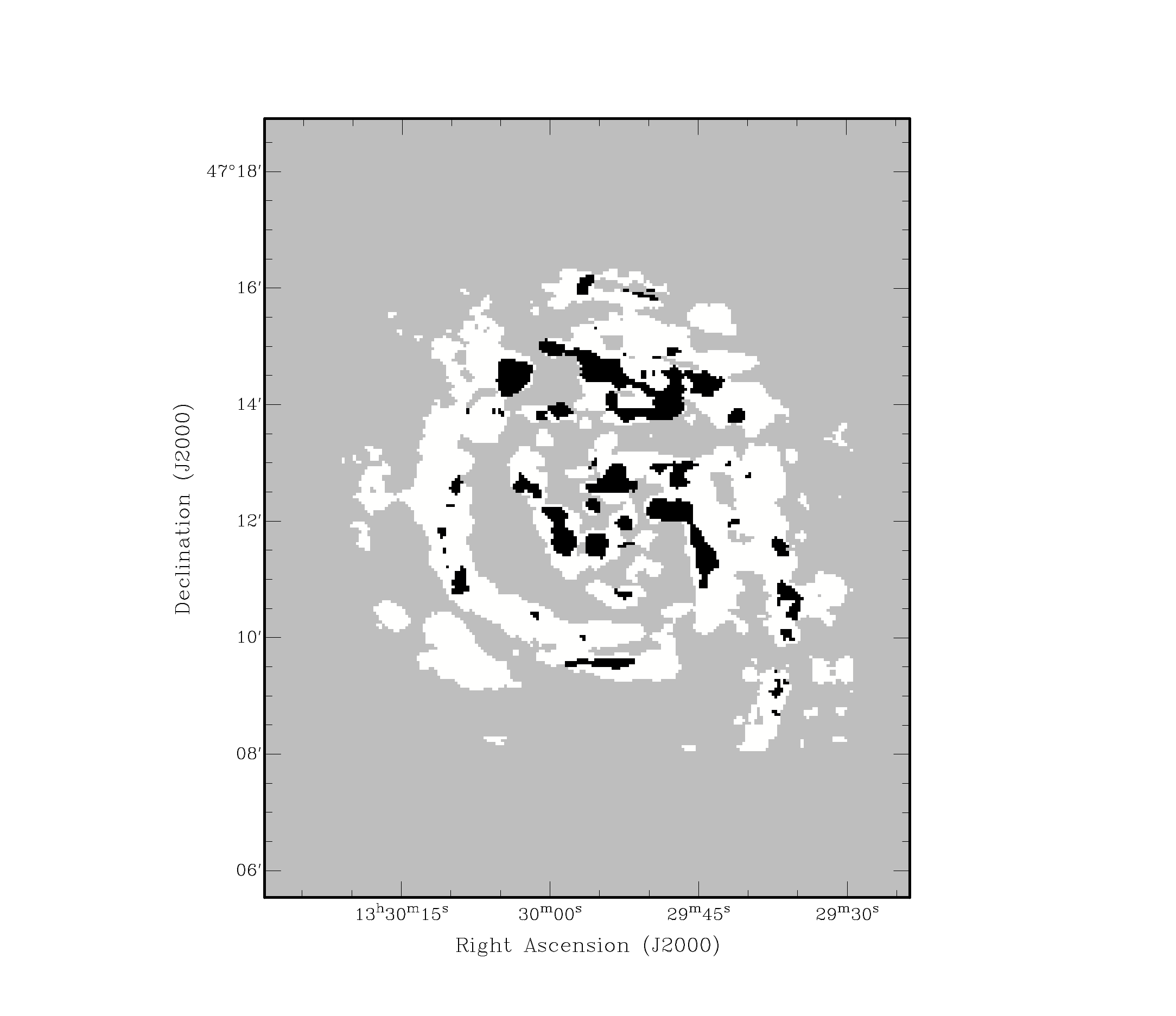}
\caption{A map of M51 showing the spatial distribution of the best-fit model from directly fitting to Stokes $Q$ and $U$. Sight lines with $Q$ $U$ versus $\lambda^2$ behaviors best fitted by a uniform external Faraday screen are denoted by the white pixels. Sight lines with $Q$ $U$ versus $\lambda^2$ behaviors best fitted by an inhomogeneous external Faraday screen are denoted by the black pixels. Pixels with insufficient signal-to-noise detection in polarization have been blanked (grey).}
\label{fig:best_model}
\end{figure*}

The Stokes $Q$ and $U$ versus $\lambda^2$ trends can be well fitted by Faraday rotation occurring in an external screen. More specifically, approximately 84\% of the sight lines are consistent with a uniform screen, without wavelength-dependent depolarization across L band, while around 16\% of the sight lines are consistent with external Faraday dispersion produced by an inhomogeneous Faraday screen. Sight lines well described by external Faraday dispersion have $\sigma_{\rm RM}$ in the range 9 - 29 rad m$^{-2}$, well exceeding the $\sigma_{\rm RM}$ due to the Milky Way foreground at angular separations $<$ 1$^\circ$. Therefore, we attribute the depolarization to M51 instead of the Milky Way. We show the spatial distribution of the derived best-fit model in Figure~\ref{fig:best_model}. We will discuss implications of this result in the next section. Since the best fit models of the medium do not cause any changes to the value of Faraday depth along each sight line, we continue to use the Faraday depth map derived using RM synthesis (\S~\ref{section:rmsynthesis_diff_emission}) in the rest of this paper.

\section{The Nature of the Faraday Rotating Medium of M51 at 1-2 GHz}
\label{section:nature_faraday_medium}

\subsection{Relative Scale Heights of Thermal and Cosmic Ray Electrons}
In Section~\ref{subsection:diffuse_qu_fit}, we showed that the observed trends of Stokes $Q$ and $U$ against $\lambda^2$ can be well fitted by models where Faraday rotation takes place outside of the synchrotron emitting region. This result on the geometry of the Faraday rotating and synchrotron emitting medium immediately implies that the scale height of the thermal gas (the warm ionized medium) is greater than that of the synchrotron emitting gas at 1.4 GHz. Statistically speaking, scale heights of these two constituents of the ISM are found to be similar: the typical synchrotron scale height of edge-on galaxies is 1.8 kpc at 6 cm \citep{krause2009}, while the mean H$\alpha$ scale height ($\propto$ $n_e^2$) is in the range of 1$-$2 kpc \citep{rossa2003}. Non-zero Faraday rotation in the halo of the Large Magellanic Cloud implies comparable thermal and synchrotron scale heights \citep{mao2012b}. An extreme case would be M31, which has a warm ionized medium three times thicker than the synchrotron emitting medium \citep{fletcher2004}. At the location of the Sun in our own Milky Way, the warm ionized medium also has a larger scale height \citep[1.8 kpc,][]{gaensler2008} than the synchrotron thick disk at 1.4 GHz \citep[$<$ 1.5 kpc,][]{beuermann1985}. A strong plane-parallel magnetic field compared to the vertical component could confine relativistic electrons closer to the mid-plane where they have been accelerated, resulting in a smaller synchrotron scale height. Since there is no observational evidence for a global outflow or a galactic-scale wind in M51, we consider only diffusion and streaming as the dominant cosmic ray transport mechanisms in this galaxy. We can place a lower limit on the  thermal electron scale height by computing the synchrotron scale height at L band using the synchrotron cooling time and the particle transport velocity. The synchrotron cooling time at frequency $\nu_{\rm GHz}$ is given by
\begin{equation}
\tau_{syn} \approx 10^9 B_{\rm \mu G}^{-3/2} \nu_{\rm GHz}^{-1/2}~ {\rm yrs}
\end{equation} 
For a total plane-of-the-sky magnetic field strength of $\sim$ 15 $\mu$G at 1.5 GHz \citep{fletcher2011}, the life time of cosmic ray electron is 1.4$\times$10$^7$ yrs. 
If we assume that the cosmic ray propagation is diffusive and a diffusion coefficient $D$ of 3$\times$10$^{28}$ cm$^2$ s$^{-1}$ as in the Milky Way \citep[e.g.][]{strong2007}, then within the lifetime of these cosmic ray electrons, they can diffuse up to a height of  $\sqrt{D\tau_{syn}}$ $\sim$ 1.2 kpc. In addition, we can place an upper limit on the height of the synchrotron disk by assuming that cosmic rays stream along field lines at the Alfv\'{e}n speed into the halo:
\begin{equation}
v_A =2\times 10^5 \frac{B_{\rm \mu G}}{\sqrt{n_i}} {\rm cm~s^{-1}}
\end{equation} 
For an ion number density of n$_i$ $\sim$ 0.1 cm$^{-3}$, the Alfv\'{e}n speed is 95 km s$^{-1}$. Within the life time of these cosmic ray electrons, they can travel up to a height of approximately 1.4 kpc from the mid-plane if the field lines are completely vertically directed.  Since the thermal gas has a larger extent than the cosmic ray electrons, the scale height of the thermal gas must exceed 1.2 kpc. We suggest that a systematic wide-band polarization survey and careful modeling of Stokes $Q$ and $U$ against $\lambda^2$ at L band of nearby low inclination galaxies would be a novel approach to extract relative scale heights of different phases of the ISM, hence allowing for a better understanding of the source of ionization of the warm ionized medium (WIM) and cosmic ray confinement in galaxies. 

\subsection{External Faraday Dispersion in the halo of M51}

\begin{figure*}
\centering
\includegraphics[width=1.0\textwidth]{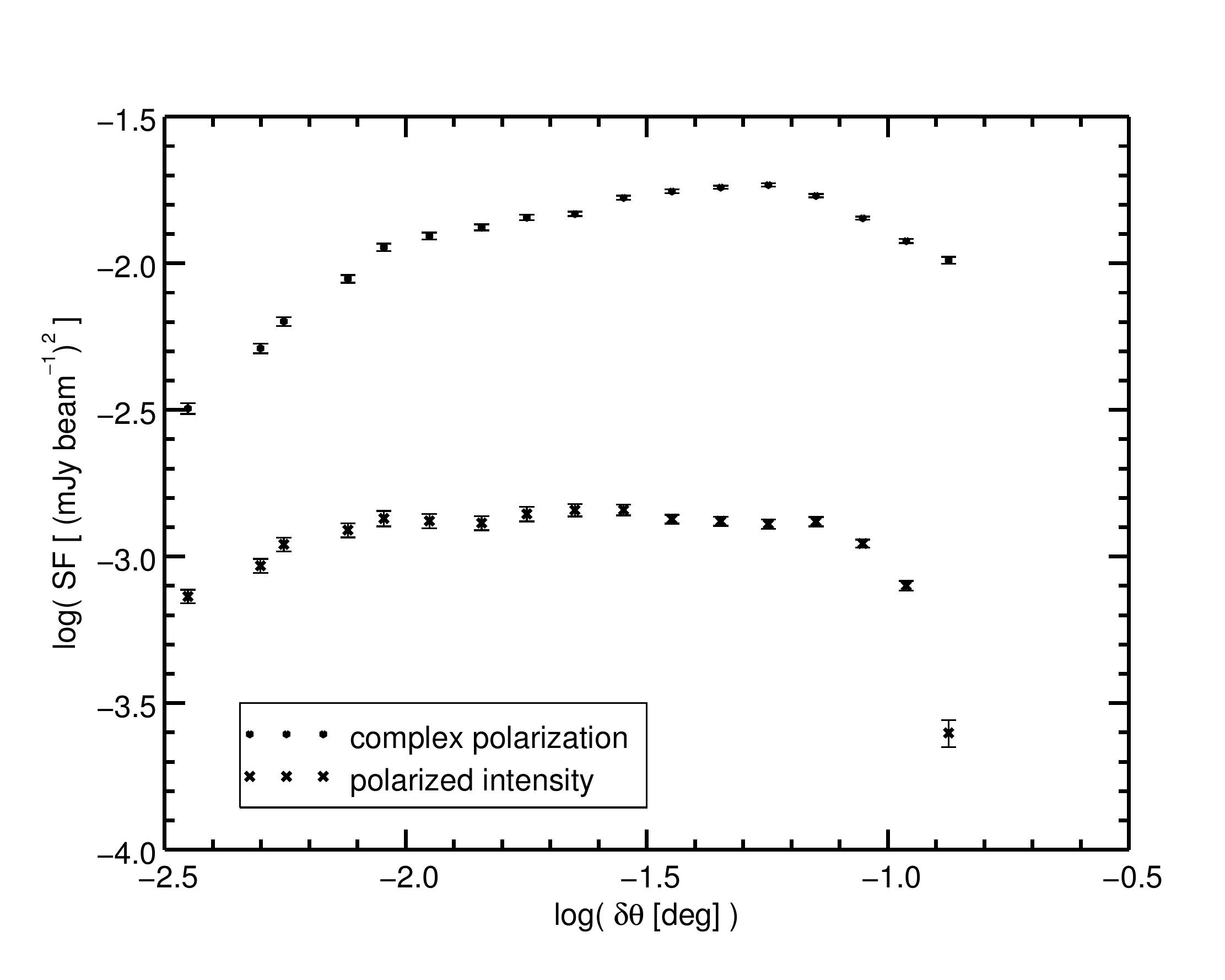}
\caption{Structure function of complex polarization (dots) and polarized intensity (crosses) across M51 at L band.}
\label{fig:sf_cp_pi}
\end{figure*}

Most sight lines towards M51 do not exhibit any $\lambda$-dependent depolarization across L band, suggesting that the external Faraday dispersion effect is not severe across M51 in this wavelength range. The Faraday screen is either completely homogeneous, or that our observations have a resolution much smaller than the characteristic turbulent scale in the screen, leading to little / no external Faraday dispersion effects. We show, in Section~\ref{section:sf_analysis}, using RM structure function analysis that the latter is indeed the case. We constructed the structure function of the complex polarization ($SF_{CP}$) and that of the polarized intensity ($SF_{pi}$) 
\begin{equation}
SF_{CP}(\delta\theta) = \langle |P(\theta) - P(\theta+\delta\theta)|^2 \rangle
\end{equation}
\begin{equation}
SF_{pi}(\delta\theta) = \langle (p(\theta) - p(\theta+\delta\theta))^2 \rangle, 
\end{equation}
where $P$ = $Q+iU$ and $p$ = $|P|$. According to \cite{sun2014a}, the relative steepness of $SF_{CP}(\delta\theta)$ and $SF_{pi}(\delta\theta)$ can be used to diagnose whether the observed polarized emission is intrinsic or heavily modified by a foreground turbulent Faraday screen which randomizes the distribution of polarization angles, reduces the observed polarized intensity and moves power to smaller scales. The authors showed mathematically that if the observed polarized structure is mainly produced by a foreground Faraday screen, then the slope of $SF_{CP}(\delta\theta)$ would be significantly shallower than $SF_{pi}(\delta\theta)$. Intuitively, this can be understood as $SF_{CP}(\delta\theta)$ being dependent on the distributions of both the intensity and the angle, and hence the slope of $SF_{CP}(\delta\theta)$ would be flatter than  $SF_{pi}(\delta\theta)$ if external Faraday dispersion is in action. In Figure~\ref{fig:sf_cp_pi}, we plot  the structure functions $SF_{CP}$ and $SF_{pi}$ computed using our L band M51 data. It is clear that $SF_{CP}(\delta\theta)$ does not have a shallower slope than $SF_{pi}(\delta\theta)$. This suggests that the observed polarized structures are mostly intrinsic with little modification/depolarization due to the foreground Faraday screen in M51's halo, which is consistent with the result of our direct $Q$$U$ fit in \S~\ref{subsection:diffuse_qu_fit}.

\subsection{Small-scale RM Fluctuation in the Halo of M51}

While most of the sight lines through M51 do not show $\lambda$-dependent depolarization within L band, 16\% of the sight lines through the galaxy do exhibit $\lambda$-dependent depolarization described by external Faraday dispersion ($\propto e^{-2\sigma_{\rm RM}^2 \lambda^4}$). Several regions of M51, including the nucleus, the inner spiral arms and the region between M51 and the companion galaxy appear to experience more severe  $\lambda$-dependent depolarization: the dominant characteristic turbulent cell size in these regions must be smaller compared to the physical scale encompassed by the beam (370 pc). Depolarization in these regions can be well described by a median RM variance ($\sigma_{\rm RM}$) of 15 rad m$^{-2}$. By evoking a simple single-cell random magnetic field model\footnotemark[7]\footnotetext[7]{Magnetic field strength in each individual cell has the same magnitude but different orientation.} \cite[e.g.][]{gaensler2001}, one can relate the random magnetic field strength in the halo $B_{r,halo}$ to the variance of RM ($\sigma_{RM}$), electron density ($n_e$), cell size ($l$), filling factor ($f$) and total path length $D$: 
\begin{equation}
B_{r,halo}=\frac{\sigma_{RM}}{0.812 n_e}\sqrt{\frac{3f}{Dl}}. 
\end{equation}
Assuming $n_e$ = 0.05 cm$^{-3}$, $l$ $<$ 370 pc, $D$ = 1 kpc and $f$ = 1, we estimate a lower limit of the random field strength in the halo $B_{r,halo}$ to be 1 $\mu$G. 

We search for spatial correlations of these more turbulent regions in M51 with properties of the underlying star-forming disk. HII regions on pc scales are a viable source of energy input into the magneto-ionic medium and they have been shown to drive RM fluctuations in the Milky Way spiral arms \citep[e.g.,][]{haverkorn2006}. A large concentration of HII regions within the telescope beam can effectively depolarize background emission. While we find some correspondence between locations of HII regions \citep{lee2011} and pixels that exhibit external Faraday dispersion within $\sim$ 3 kpc from the nucleus, there is little correspondence in the rest of the galaxy. This is expected, since we have argued that polarized emission at L band likely originates from the top of the synchrotron emitting disk, there should be little or no correlation of depolarization structures with HII regions in the underlying disk. 

We further investigate whether the more turbulent regions revealed by polarimetry correspond to more turbulent regions in neutral hydrogen. We use HI velocity dispersion of M51 from the THINGS survey \citep{walter2008} and find that the sight lines exhibiting $\lambda$-dependent depolarization within L band appear to have a slightly higher mean velocity dispersion (24.0$\pm$0.2 km s$^{-1}$) than the rest of the galaxy (21.60$\pm$0.08 km s$^{-1}$). We note that HI velocity dispersion and depolarization of synchrotron emission are completely different approaches that probe turbulence in two different phases of the ISM. The fact that these two very different measurements converge suggests that the general ISM near M51's nucleus and in the region between M51 and the companion NGC 5195 has been stirred up. Modeling depolarization trends of synchrotron emission is thus an alternative way to identify and characterize turbulent regions in a galaxy. 

\subsection{Reconciling the Derived Faraday Medium Structure with Low Frequency Observations of M51}

We check the consistency of our derived Faraday structure in M51's halo with low frequency polarization observations of the galaxy, specifically the recent Giant Metrewave Radio Telescope (GMRT) observations at 610 MHz \citep{farnes2013}. M51 is found to be completely depolarized down to a sensitivity limit of 44 $\mu$Jy beam$^{-1}$ at a resolution of 24" with a bandwidth of 16 MHz. We repeated our direct $Q$ $U$ fit (Section~\ref{subsection:diffuse_qu_fit}) after smoothing the input Stokes $Q$ and $U$ cubes to a resolution of 24". Assuming a synchrotron spectral index of $-$1.1 and accepting a more complex model if the F-test gives a significance level $>$ 4$\sigma$, the predicted median polarized intensity of M51 at 610 MHz is 30 $\mu$Jy beam$^{-1}$, below the sensitivity limit of the GMRT observations. Thus, the non-detection of M51 in polarization at 610 MHz is consistent with our model at the 4 $\sigma$ level.
Since cosmic ray electrons suffer less from energy losses and can propagate further away from their acceleration sites at low frequencies, the synchrotron scale height is likely larger at lower frequencies. For example, if cosmic ray diffusion dominates, then the synchrotron scale height should vary as $\nu^{-1/4}$. We suggest that at frequencies below 1 GHz, a part of the near-side halo may emit as well as Faraday rotate. Since this emission will suffer from differential Faraday rotation and internal Faraday dispersion, it is unlikely to produce a significant amount of polarization. Future deep wide-band polarization observations of M51 at low frequencies are much needed to further constrain the properties of M51's magnetized near-side halo.

\section{Large-Scale Magnetic Field Structures in the Halo of M51}
\label{section:large_scale_bfield_model}
\subsection{The Fletcher Model}

In the comprehensive modeling work of \cite{fletcher2011}, based on the drastically different Faraday depths and degree of polarization measured at long (18cm and 20cm) and short wavelengths (3cm and 6cm), the authors concluded that there must be two distinct layers where Faraday rotation occur in M51, designated as the disk and the halo. In particular, polarized emission at 18 and 20cm originates from the top of the synchrotron emitting disk, and only experiences Faraday rotation in the thermal halo. Therefore, the Faraday structure of the \cite{fletcher2011} model agrees well with the result of our direct $Q$$U$ fit (see Section~\ref{subsection:diffuse_qu_fit}): the thermal gas has a larger scale height than the synchrotron emitting gas at L band. 

\begin{figure*}
\centering
\begin{subfigure}
\centering
\includegraphics[width=2.3in,clip=true]{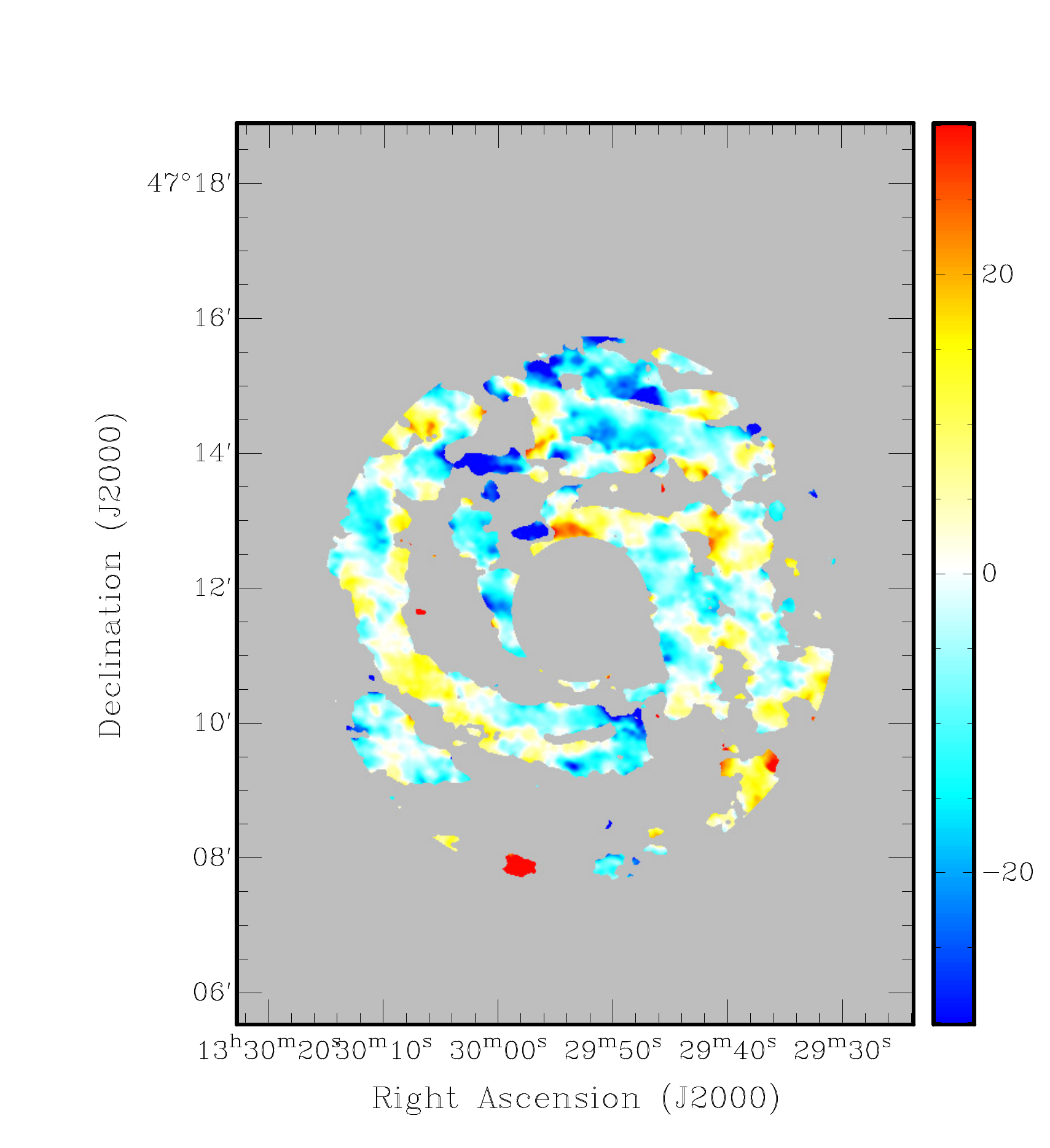}
\end{subfigure}

\begin{subfigure}
\centering
\includegraphics[width=2.3in,clip=true]{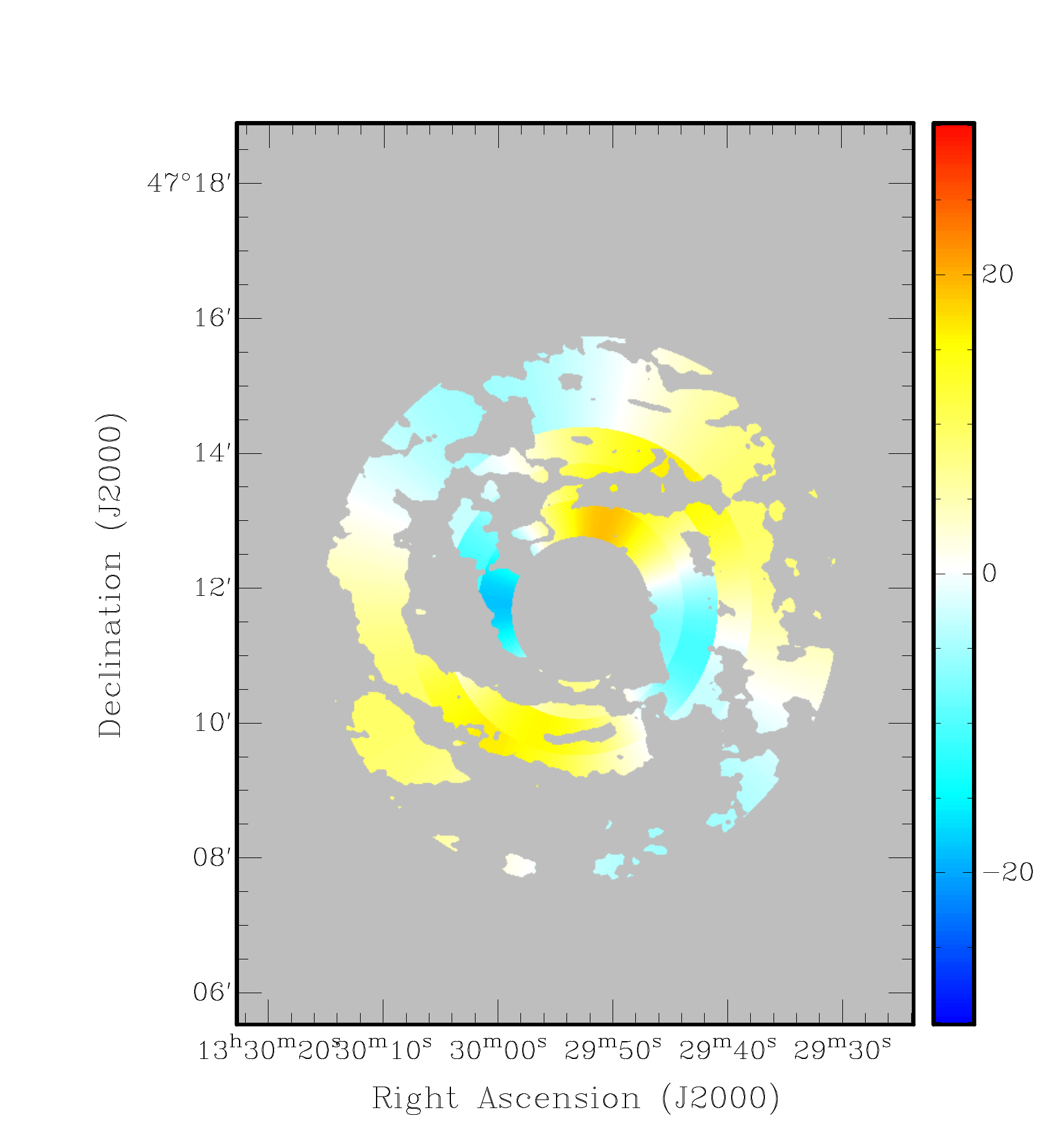}
\end{subfigure}

\begin{subfigure}
\centering
\includegraphics[width=2.3in,clip=true]{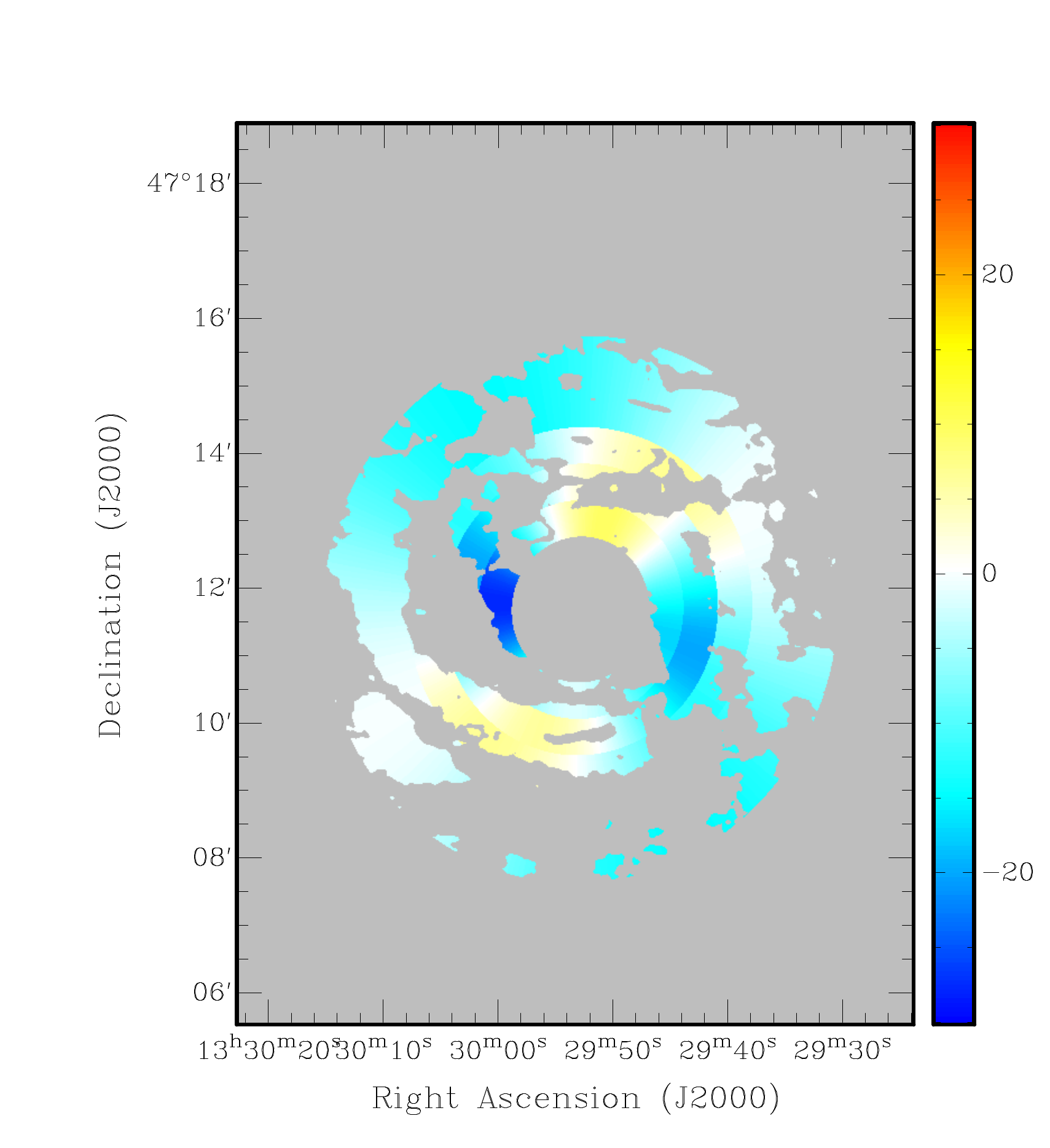}
\end{subfigure}
\caption{Top: Faraday depth distribution of M51 at L band after removing the constant Milky Way contribution of $+$13 rad m$^{-2}$. Middle : The predicted Faraday depth distribution of M51 from the \cite{fletcher2011} bisymmetric halo model. Bottom: The predicted Faraday depth distribution of M51 from the \cite{fletcher2011} bisymmetric halo field with the addition of a vertical field that produces a Faraday depth of $-$9 rad m$^{-2}$.}
\label{fig:rm_map_3panels}
\end{figure*}

Next, we compare the observed Faraday depths at L band with predictions from the \cite{fletcher2011} model. By fitting Fourier modes to the observed polarization angles averaged in azimuth sectors at four sparsely sampled wavelengths, \cite{fletcher2011} found that M51 hosts disk and halo fields of different topologies: the disk field is axisymmetric (with a weak m=2 component), while the halo field is bisymmetric. In the top panel of Figure~\ref{fig:rm_map_3panels}, we show the observed Faraday depth of M51 after removing the Milky Way foreground RM (+13 rad m$^{-2}$). In the middle panel of the same figure, we show the predicted Faraday depth distribution of M51 at L band produced by the halo bisymmetric field in the Fletcher model\footnotemark[8]\footnotetext[8]{We have extended the model prediction out to a galacto-centric radius of 9 kpc.}. It is clear that the \cite{fletcher2011} prediction is on average much too positive compared to the observed values. 

The RM due to the Milky Way foreground is a free parameter in the Fletcher model that is found to be $\sim$ +4 rad m$^{-2}$ across all radii. This value is different from our estimated Milky Way RM based on the median Faraday depths of background extragalactic sources in the field (+13 rad m$^{-2}$). There exists a degeneracy between the Milky Way RM and an RM produced by a vertical magnetic field in the halo of M51. Like the Milky way foreground, a vertical magnetic field in the halo of M51 will produce additional Faraday rotation at both short (3 and 6 cm) and long (18 and 20 cm) wavelengths. In the bottom panel of Figure~\ref{fig:rm_map_3panels}, we show the predicted Faraday depths of M51 at L band from the \cite{fletcher2011} model with the addition of a vertical field that produces an RM of $-$9 rad m$^{-2}$. This map clearly resembles the observed Faraday depth distribution much better -- the value of the summed residual  $\Sigma_i$ (RM$_{\rm model,i}$$-$RM$_{\rm observed,i}$)$^2$ decreases significantly from 1.6$\times$10$^7$ to 1.3$\times$10$^7$. We suggest that the discrepancy between the observed Faraday depths of M51 at L band and the Fletcher model prediction can be reconciled if we include a coherent vertical magnetic field in M51's halo in addition to the bisymmetric plane-parallel component\footnotemark[9]\footnotetext[9]{This coherent vertical field has a small component projected onto the sky plane but this component does not emit synchrotron radiation due to the lack of cosmic ray electrons in the halo.}. A similarly weak but coherent vertical magnetic field is also present in the halo of the nearby Large Magellanic Cloud \citep{mao2012b} and M33 \citep{tabatabaei2008}.

This coherent vertical field could be of either primordial or of dynamo origin. While azimuthal fields in the disk of the galaxy can diffuse away easily, the vertical magnetic flux is essentially conserved \footnotemark[10]\footnotetext[10]{The vertical magnetic flux is conserved unless there is a radial flow that drags field lines out of the galaxy.}\citep[e.g.][]{sofue1987}. Thus, the measured vertical field in the halo of M51 could reflect the field preserved since the collapse of the protogalaxy. The expected magnitude of the vertical field in this scenario depends heavily on the initial seed field strength. On the other hand, dynamo theory predicts vertical fields $\sim$ 10\% of the plane-parallel component. In particular, the ratio of vertical field to radial field follows from dynamo theory is given by $\sqrt{h/R}$ \citep{ruzmaikin1988} where $h$ is the disk scale height and $R$ is the radius of the galaxy. Assuming $h$$\sim$ 1 kpc, $R$ $\sim$ 10 kpc for M51 and the derived radial component of the magnetic field in the \cite{fletcher2011} model, we obtain $|$RM$_{\rm{B_z}}$$|$ in the range of 7$-$21 rad m$^{-2}$ in the halo. Our derived magnitude of $\sim$ 9 rad m$^{-2}$ falls comfortably in this range. 
\subsection{The Braun Model}
\subsubsection{Large-Scale Azimuthal Trends of Faraday Depth and Polarized Intensity}

Using a sample of WSRT SINGS galaxies, \cite{braun2010} proposed that galaxies have axisymmetric spiral fields with a quadrupole field geometry in the nearside halo. At L band, the polarized emitting zone is concentrated 2-4 kpc above the mid-plane, this emission is then Faraday rotated at a height of 4$-$10 kpc above the mid-plane in the near-side halo. This picture of Faraday rotation taking place in a purely thermal halo is consistent with the result of our direct $Q$$U$ fit in Section~\ref{subsection:diffuse_qu_fit}, though the exact vertical extents of the emitting and rotating zones may not agree.

\begin{figure*}[h]
\centering
\includegraphics[width=1.0\textwidth]{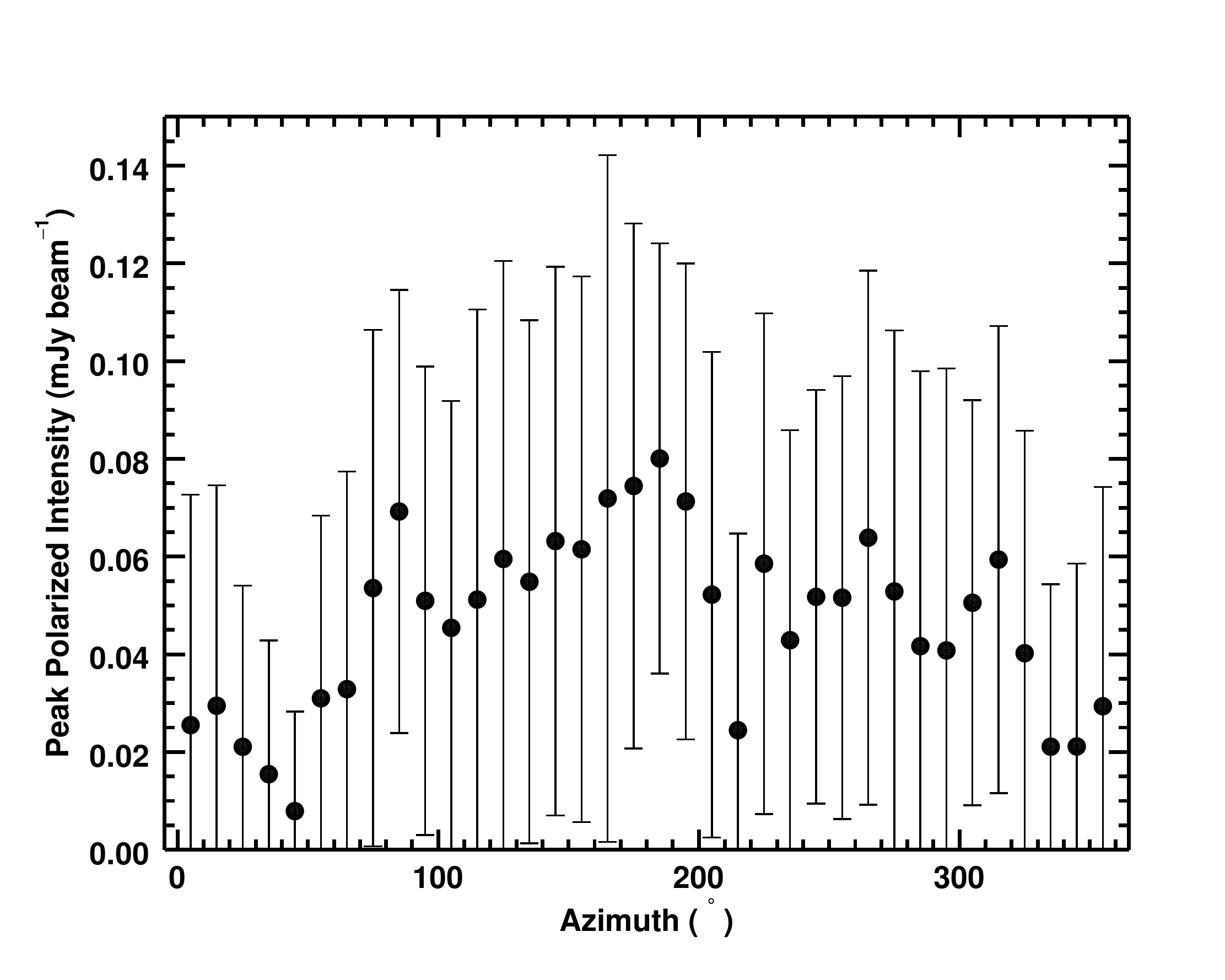}
\caption{Peak polarized intensity in our Faraday depth cube plotted as a function of azimuth. Only pixels with signal-to-noise detection in polarization greater than 7 have been averaged in 10$^\circ$ bins across all radii. Error bars represent the standard deviation of polarized intensity within a bin.}
\label{fig:mao_pi_vs_az}
\end{figure*}

\begin{figure*}
\centering
\includegraphics[width=1.0\textwidth]{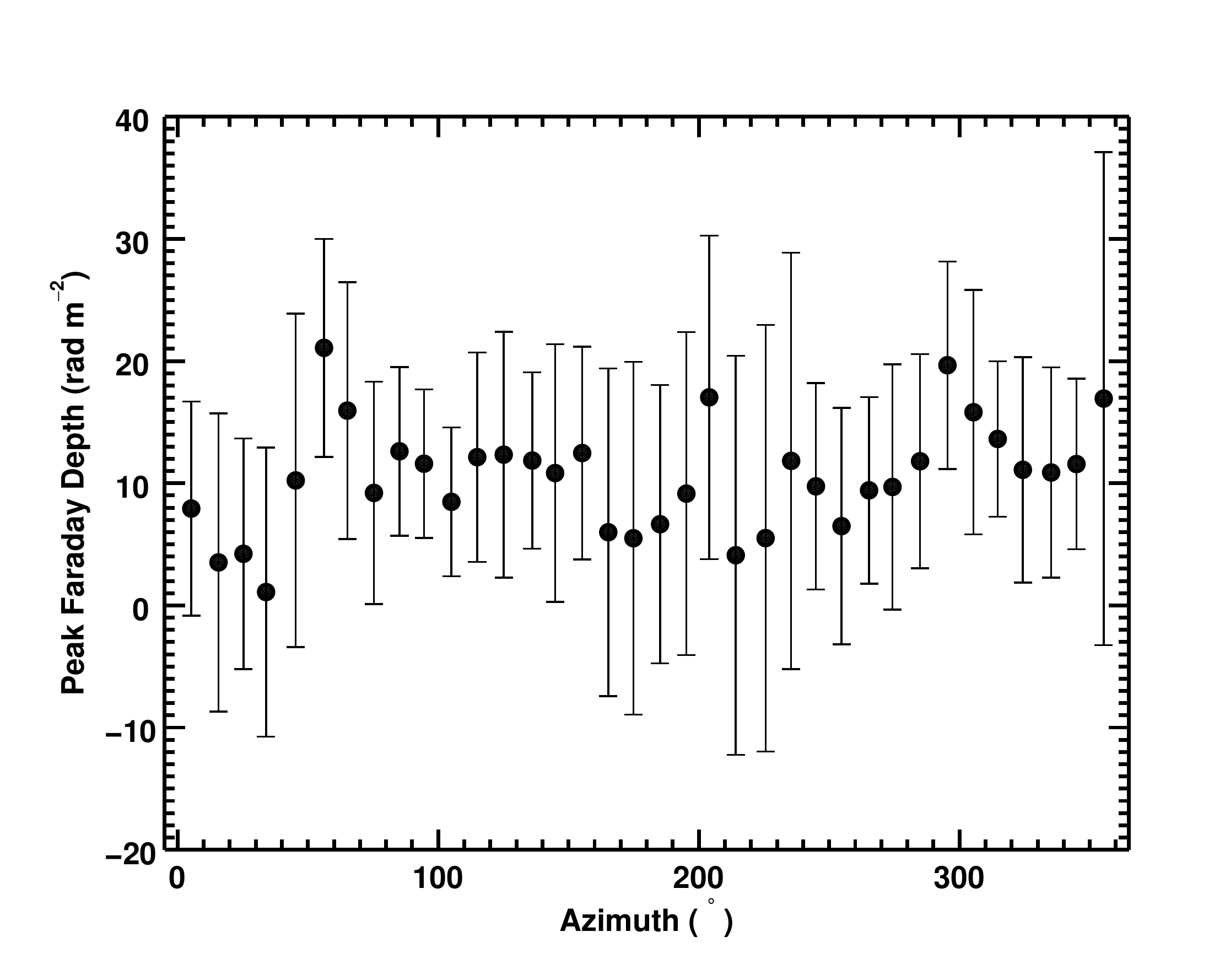}
\caption{Peak Faraday depth in our Faraday depth cube plotted as a function of azimuth. Only pixels with signal-to-noise detection in polarization greater than 7 have been averaged in 10$^\circ$ bins across all radii. Error bars represent the standard deviation of Faraday depth within a bin.}
\label{fig:mao_fd_vs_az}
\end{figure*}

 The geometry of the magnetic field manifests itself in the polarized intensity versus azimuth and Faraday depth versus azimuth trends. In particular, one expects asymmetry in polarized emission at L band with a global minimum in polarized intensity near the receding major axis\footnotemark[11]\footnotetext[11]{For M51, the position angle of the receding major axis is 170$^\circ$ \citep{shetty2007}.}. Meanwhile, one expects a minimum Faraday depth (a value closest to the Galactic foreground) near the approaching major axis (position angle of $-$10$^\circ$). In Figure~\ref{fig:mao_pi_vs_az} and \ref{fig:mao_fd_vs_az}, we plot the measured peak polarized intensity as a function of azimuth (AZ = 0$^\circ$ at the receding major axis) and the peak Faraday depth as a function of azimuth averaged across all radii in 10$^\circ$ bins. Error bars in the plot represent the standard deviations of polarized intensity and Faraday  depths within a bin. We can not readily identify any visible azimuthal trends in these plots. Our derived polarized intensity and Faraday depth modulations as a function of azimuth are visibly different from those presented by \cite{braun2010}. This disparity could potentially be caused by a different de-projection of the galaxy: we have adopted an inclination angle of 20$^\circ$ \citep{tully1974,shetty2007}, whereas \cite{braun2010} assumed an inclination of 42$^\circ$. The use of a different signal-to-noise cutoff in polarization detection when making the azimuth plots may also lead to different azimuthal modulations. 
 
In the \cite{braun2010} work, instead of directly predicting the azimuth modulations of the Faraday depth and polarized intensity, trends of the integrated line-of-sight ($\int$B$_{||}$$dl$) and plane-of-the-sky ($\int$B${_\perp}$$dl$) magnetic fields were presented. The observed Faraday depth is a convolution of the line-of-sight magnetic field and the thermal electron density distribution while the observed polarization is a convolution of the plane-of-the-sky magnetic field and the cosmic ray electron density distribution. Unless the thermal electron and cosmic ray electron distributions are completely uniform across the entire disk, trends of Faraday depth and polarized intensity may not reflect those of $\int$B$_{||}$$dl$ and $\int$B${_\perp}$$dl$. Moreover, the predicted modulations from their magnetic field model were plotted at a single fixed galacto-centric radius, but not averaged across all radii as how the observed trends were presented. It is unclear whether these characteristic modulations persist when Faraday depth and polarization averaging has been performed across all radii within an azimuth bin. In order to quantitatively compare the \cite{braun2010} model and our observed Faraday depth and polarized intensity trends, a realistic diffuse ionized gas and cosmic ray electron distribution of the modeled galaxy is needed so that trends of Faraday depth and polarized intensity as a function of azimuth can be predicted directly and compared with the observed values. This topic is outside of the scope of the current paper and will be addressed in our forthcoming work. 
\subsubsection{A Search for Polarized Emission from the Far-side Halo}
\label{subsection:backside_halo}
\begin{figure*}
\centering
\includegraphics[width=0.5\textwidth]{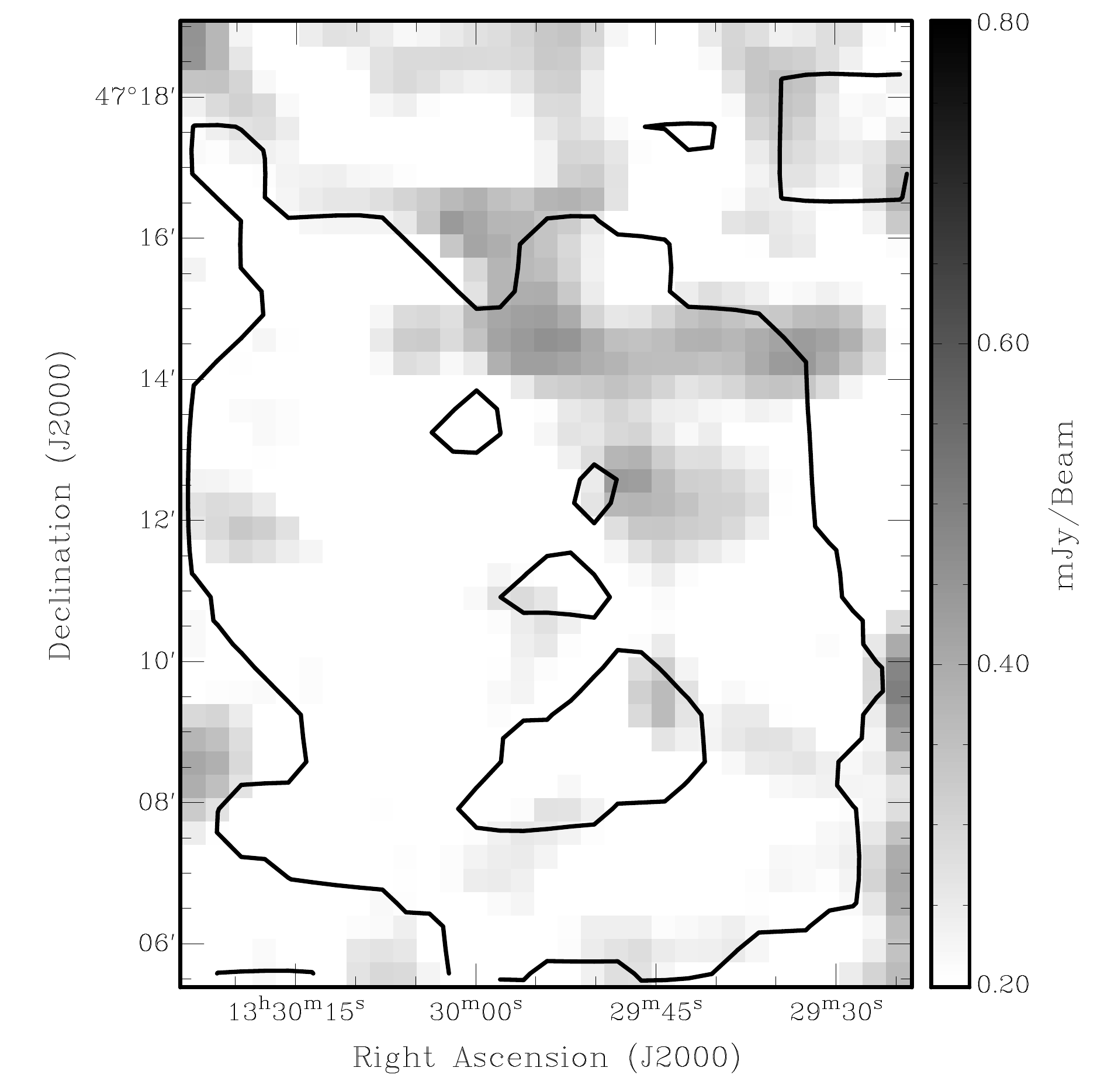}
\includegraphics[width=0.5\textwidth]{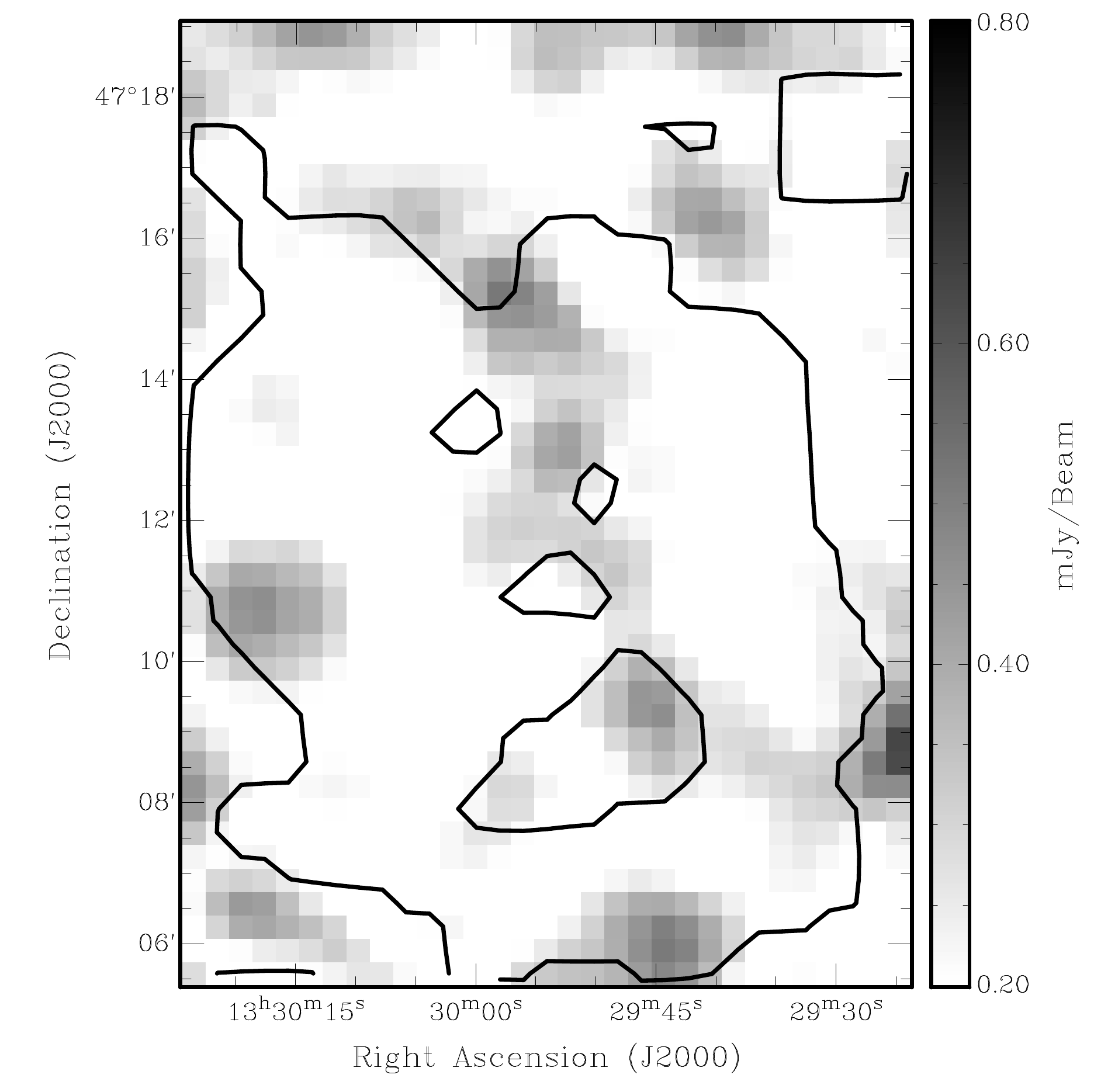}
\caption{Polarized emission of M51 at a Faraday depth of $-$180 rad m$^{-2}$ (left) and $+$200 rad m$^{-2}$ (right) at a resolution of 90". The contour roughly outlines the location of M51: it is the polarized intensity contour at the peak of the Faraday depth cube at a level of 0.96 mJy beam$^{-1}$. No significant polarized emission is detected.}
\label{fig:far_side_halo}
\end{figure*}

\cite{braun2010} proposed that polarized emission of a sample of SINGS galaxies at cm wavelengths arises from a zone 2-4 kpc above the mid-plane in the near-side halo. If the galaxy is symmetric, then there should exist a similar synchrotron emitting zone 2-4 kpc below the mid-plane. \cite{braun2010} found evidence for this far-side halo in their WSRT Faraday cube of M51 at a level of 0.6 mJy beam$^{-1}$ at a resolution of 90". Since this polarized emission from the far-side halo must first propagate through the galactic disk before reaching us, it will be dispersed to Faraday depths of $\sim$ $\pm$200 rad m$^{-2}$ due to the disk magnetic field and it will also be depolarized (by external Faraday dispersion effects) due to strong fluctuations of magnetic fields and electron densities in the turbulent disk. We search for the polarized emission from the far-side halo in our new Jansky VLA L band data of M51. To increase our sensitivity to weak and extended emission, we smooth the input Stokes $Q$ and $U$ image cubes to a resolution of 90". We then proceed by remaking the Faraday depth cube at this coarser resolution. Noise in the resulting Faraday depth cube is $\sim$ 0.12 mJy beam$^{-1}$. Hence, if there exists polarized emission $>$ 0.6 mJy beam$^{-1}$, it should be easily detected at a signal-to-noise level above 5. In Figure~\ref{fig:far_side_halo}, we show the polarized emission of M51 at Faraday depths of $-$180 rad m$^{-2}$ and $+$200 rad m$^{-2}$. 

Contrary to the findings of \cite{braun2010}, we do not detect significant polarized emission at these Faraday depths. There are several possible reasons for this non-detection. First, the polarized emission detected by \cite{braun2010} may be associated with the first side-lobes of the WSRT RMSF, as the authors have also cautioned in their work. The first side-lobes of the WSRT RMSF are located at roughly the same Faraday depths as the expected emission from the far-side halo ($\pm$200 rad m$^{-2}$), and they have an amplitude $\sim$ 78\% of the main peak (see Figure~\ref{fig:compare_rmsf}). Our new Jansky VLA data cover a more continuous frequency range and have suppressed the first side lobes amplitude down to $<$ 35\% of the main peak. This makes it much more difficult to confuse side-lobes with real emission when deconvolving the dirty Faraday depth spectra. Secondly, we have performed spatial smoothing in the input $Q$ $U$ cubes (i.e. in $\lambda^2$ domain) while \cite{braun2010} smoothed the complex polarization in Faraday depth space. Since RMCLEAN is a non-linear operation, artifacts can arise when spatial smoothing is done to the deconvolved Faraday depth cube.

If the far-side halo indeed exists, our non-detection can be used to estimate turbulent properties in the galactic mid-plane. We assume that the intrinsic brightness of the far-side halo is the same as that of the near-side halo (median value $\sim$ 1.2 mJy beam$^{-1}$ at 90" resolution). From our Faraday depth cube at 90" resolution, we find a 3$\sigma$ upper limit of the emission from the far-side halo to be 0.36 mJy beam$^{-1}$. We further assume that the turbulent mid-plane acts as an external inhomogeneous Faraday screen that depolarizes emission from the far-side halo according to P/P$_0$ = e$^{-2\sigma_{\rm RM}^2\lambda^4}$. This leads to a lower limit estimation for the RM variance $\sigma_{\rm RM}$ $\ge$ 20 rad m$^{-2}$ within the 90" beam. We note that this is consistent with the RM variance inferred from the structure function analysis using 3cm and 6cm RM data (Section~ \ref{section:structure_function_36cm}).

\section{Rotation Measure Structure Function Analysis}
\label{section:sf_analysis}

Structure function of rotation measure contains unique information on the structure in the magneto-ionic medium, including turbulence. The second order structure function is defined as 
\begin{equation}
SF_{\rm RM}(\delta\theta) = \langle [{\rm RM}(\theta) - {\rm RM}(\theta+\delta\theta)]^2 \rangle~,
\end{equation}
where $\theta$ is the projected angular separation between two sight lines and the angular brackets denote the average of independent measurements having the same range of angular separation $\delta\theta$. Uncertainties in the RM measurements contribute to the observed RM structure function in the form of a DC offset $\langle \delta {\rm RM}(\theta)^2 +\delta {\rm RM}(\theta+\delta\theta)^2 \rangle$, which is subtracted from the measured $SF_{\rm RM}(\delta\theta)$.

Obtaining robust uncertainties associated with $SF_{\rm RM}(\delta\theta)$ is essential to correctly interpret the structure function. Since $[{\rm RM}(\theta) - \rm{RM}(\theta+\delta\theta)]^2$ is a positive-only quantity, values within a $\delta\theta$ bin are not Gaussianly distributed. Therefore, Gaussian statistical indicators of the scatter (standard deviation and standard error in the mean) may no longer be good representations of the true scatter within a $\delta\theta$ bin. To obtain realistic error estimates, we utilize the bootstrap method. While the regular bootstrap approach will spatially de-correlate Faraday rotation between sight lines and erase all features in $SF_{\rm RM}(\delta\theta)$, the marked point bootstrap method is well suited for spatially correlated data\footnotemark[12]\footnotetext[12]{This variant of the traditional bootstrap method has been used to estimate errors of the two-point correlation function of galaxy distribution.}: instead of resampling the individual RM values, we resample $[{\rm RM}(\theta) - {\rm RM}(\theta+\delta\theta)]^2$  in order to preserve spatial correlations \citep{loh2008}. The original RM sample is bootstrapped 500 times and the standard deviation of the computed $SF_{\rm RM_{\rm resampled}}(\delta\theta)$ in each $\delta\theta$ bin provides a good estimate for the error in $SF_{\rm RM}(\delta\theta)$. Throughout this analysis, we assume that the Milky Way foreground towards M51 is constant ($+$ 13 rad m$^{-2}$), and hence it does not contribute to the structure function. Any features seen in $SF_{\rm RM}(\delta\theta)$ must therefore be intrinsic to M51. 

Since RM is the line-of-sight integral of the magnetic field weighted by the thermal electron density, features in the RM structure function could reflect structures in electron densities, magnetic fields and/or the path length. In the following analysis, we assume that the polarized emission at a given wavelength (3/6 cm and L band) across M51 emerges from the same physical depth in the galaxy: for the 3/6cm data through the entire galaxy, and for the L band data from the top of the synchrotron disk. Hence, structure function features reflect only structures in electron densities and magnetic fields but not the path length difference. Since both the short wavelength (3,6 cm) and long wavelength (L band) RM maps exhibit sign reversals, we are confident that fluctuations in magnetic fields must be present and must contribute to the observed RM structure functions. To disentangle which features in the RM structure function are due to the thermal electron distribution or magnetic field structures requires independent information on the path-length integrated electron density distribution. This is provided by emission measure (EM) structure functions using extinction-corrected H$\alpha$ data or by pulsar dispersion measure (DM) structure functions (only possible for Milky Way works). However, a joint interpretation is often challenging since the actual path length probed by EM, DM and RM is likely to be very different.

\subsection{Turbulence in the Disk: RM Structure Function Using 3 and 6 cm Data}
\label{section:structure_function_36cm}

\begin{figure*}
\centering
\includegraphics[width=1.0\textwidth]{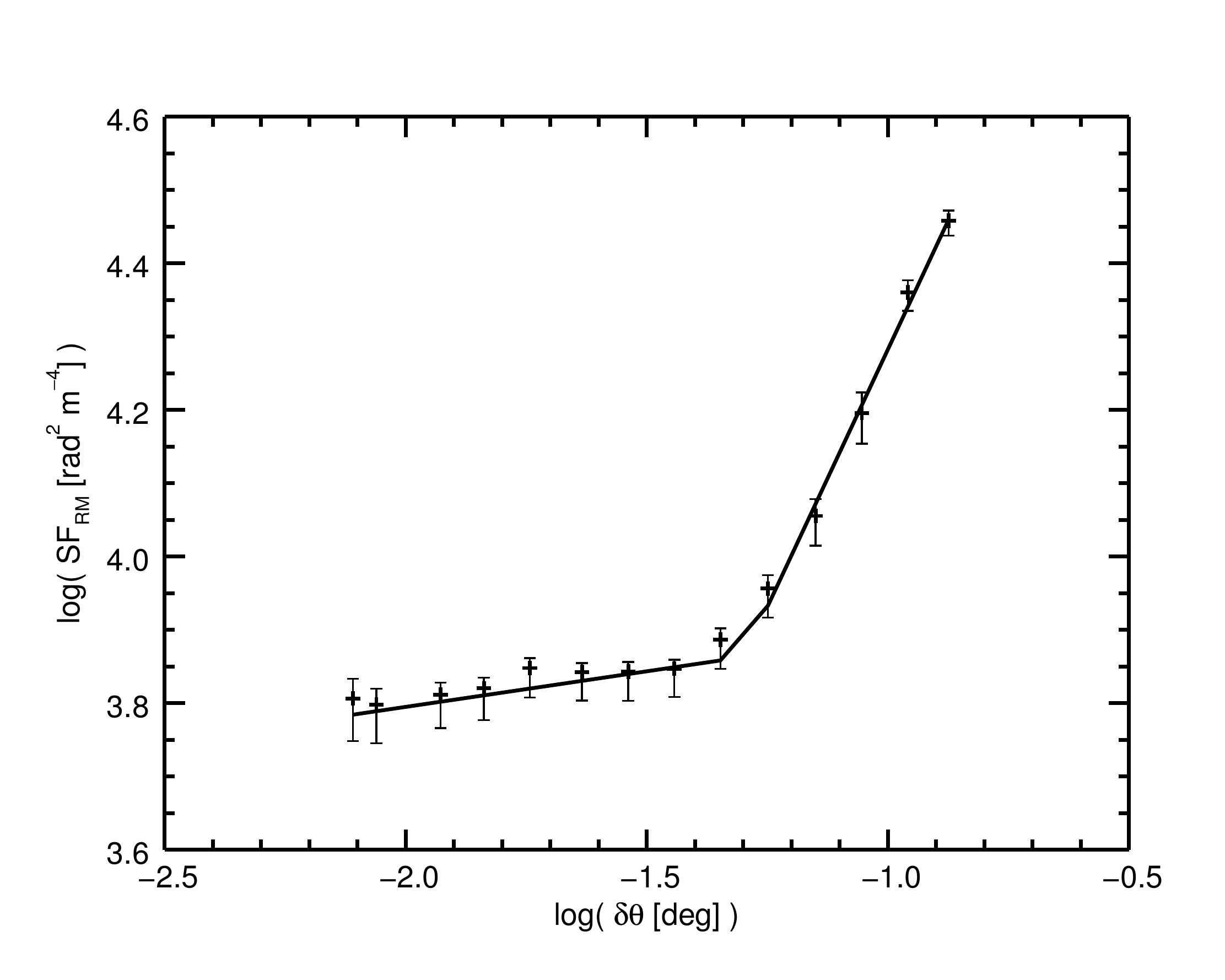}
\caption{Rotation measure structure function of M51 constructed from the short wavelength (3 and 6 cm) RM map. The structure function has been binned in equal log interval of 0.1. The error bars denote the bootstrapped uncertainty of SF$_{\rm RM}$ in each bin. The solid line is a broken power law fit to SF$_{\rm RM}$ with a break at 3'.}
\label{fig:sfrm_63cm}
\end{figure*}

We construct $SF_{\rm RM}(\delta\theta)$ using the rotation measure map derived from 3 and 6 cm VLA data by \cite{fletcher2011} which has a typical RM error of 10 rad m$^{-2}$. The resulting RM structure function is shown in Figure~\ref{fig:sfrm_63cm}. We do not display the structure function on scales smaller than the resolution of the RM map (log$\delta\theta$$<$ $-$2.3; $\delta\theta$$<$15") because any trends on these scales are affected by resolution effects. The structure function computed using short wavelength RMs can be well fitted by a broken power law with a turnover at  log$\delta\theta$ = $-$1.3 ($\delta\theta$=3', or a physical scale 6.7 kpc). Below this scale, $SF_{\rm RM}(\delta\theta)$ is flat with a slope of  0.10$\pm$0.04, similar to structure function slopes found in RM maps of diffuse polarized emission towards the inner Milky Way \citep{haverkorn2004}. On larger scales, the structure function steepens to a slope of 1.40$\pm$ 0.08.

The amplitude of the flat part of the structure function saturates at twice the RM variance 2$\sigma_{\rm RM}^2$ = 10$^{3.8}$ rad$^2$ m$^{-4}$, which corresponds to $\sigma_{\rm RM}$ = 56 rad m$^{-2}$ (since we do not spatially resolve the turbulence, the true RM variance is a factor of $\sqrt{N}$ larger, where N is the number of turbulent cells within the beam). Since 3 and 6 cm polarization data suffer much less Faraday depolarization effects, the emission likely probes through the entire galaxy. Hence, the large RM variance reflects the strong fluctuation in electron densities and magnetic fields in the turbulent mid-plane of M51. The implied $\sigma_{\rm RM}$ is in agreement with the non-detection of the far-side synchrotron emitting halo in Section~\ref{subsection:backside_halo}, from which we derived a lower limit for $\sigma_{\rm RM} \ge$ 20 rad m$^{-2}$. We note that  $\sigma_{\rm RM}$ in M51's disk is similar in magnitude to that found in the Large Magellanic Cloud  ($\sim$ 80 rad m$^{-2}$) \citep{gaensler2005,mao2012b} using structure function of polarized extragalactic sources behind the LMC. 

On scales smaller than 3', $SF_{\rm RM}(\delta\theta)$ has a much shallower slope than the expected value of 5/3 (or 2/3) for 3D (or 2D) Kolmogorov turbulence. If fluctuations in electron densities and magnetic fields in M51's disk indeed follow the Kolmogorov spectrum, then the structure function must turnover at a scale below the resolution of the current 3 and 6 cm RM map -- the outer scale of turbulence in the mid-plane of the galaxy thus must be smaller than 15", which corresponds to 555 pc at the distance of M51. This is consistent with the recent result of \cite{houde2013}, who found a turbulent correlation scale $<$ 100 pc by applying angular dispersion analysis to VLA polarization angle data at 6 cm. Based on the dispersion of Faraday rotation derived from 3 and 6 cm VLA data, \cite{fletcher2011} also independently estimated a turbulent cell size of  $\sim$ 50 pc. We note that the lack of features in $SF_{\rm RM}(\delta\theta)$ on scales $<$ 3' indicates that the random magnetic field and electron density fluctuations far dominates over any features produced by the coherent magnetic field in the disk and the halo of M51. 

To verify that the rise in the structure function on scales larger than 3' is real and not merely due to the sampling of pixels in the field of view, we scramble the RM values across the image and recompute the structure function. Since the break at 3' disappears from the structure function and $SF_{\rm RM}(\delta\theta)$ becomes completely flat, we conclude that this feature is unlikely an artifact. A large-scale RM gradient produced by coherent fields in magnetic arms of M51 could be responsible for producing this steep feature in $SF_{\rm RM}(\delta\theta)$. Similar features have also been detected in the RM structure function of NGC 6946 \citep{beck2007}, for which the authors also suggested a magnetic-arm origin. We verify that a large-scale azimuthal magnetic field in a ring-like configuration oriented at an inclination angle of 20$^\circ$ could indeed produce an RM structure function with a slope of $\sim$ 2 at large angular lags. 

\subsection{Turbulence in the Halo: RM Structure Function using L band data}
\begin{figure*}
\centering
\includegraphics[width=1.0\textwidth]{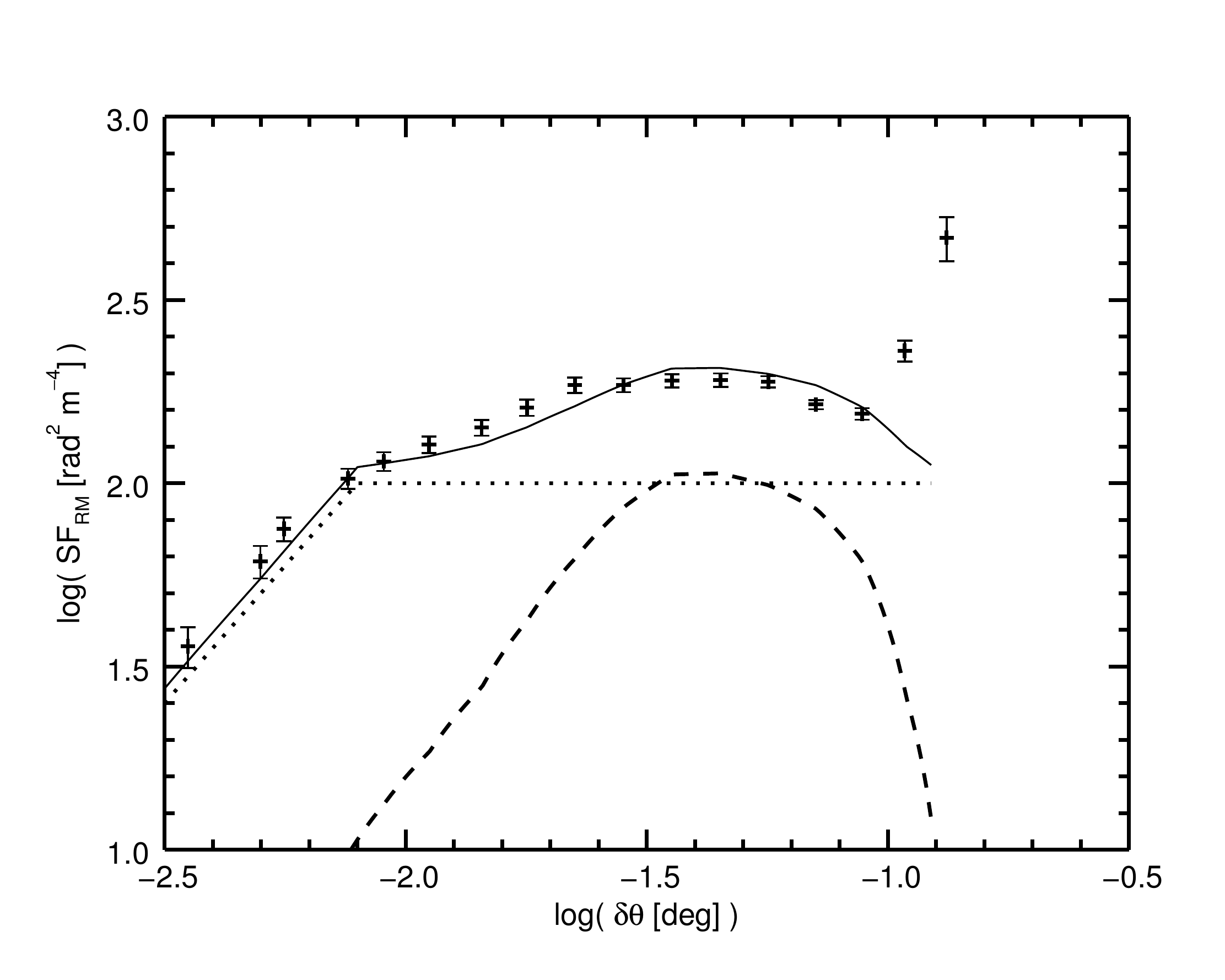}
\caption{Rotation measure structure function of M51 constructed from our newly derived L band RMs. The structure function has been binned in equal log interval of 0.1. The error bars denote the bootstrapped uncertainty of SF$_{\rm RM}$ in each bin. The dashed line represents the RM structure function expected from the best fit large-scale halo magnetic field in \cite{fletcher2011}. The dotted line represents the structure function expected from three-dimensional Kolmogorov turbulence that has an outer scale of 29" (approximately 1 kpc at the distance of M51) and a variance of 7 rad m$^{-2}$. The solid line is the sum of the structure function of the large-scale halo field and that of the Kolmogorov turbulence.}
\label{fig:sfrm_lband}
\end{figure*}

The RM structure function computed using our new Jansky VLA Faraday depth map at L band is shown in Figure~\ref{fig:sfrm_lband}. We do not display the structure function on scales smaller than the beam size (log$\delta\theta$$<$$-$2.6 or $\delta\theta$$<$11") because it is affected by resolution effects. The L band RM structure function behaves very differently from the 3 and 6 cm RM structure function, and it cannot be satisfactorily fitted by a broken power law. The general trend of the structure function is as follows: $SF_{\rm RM}(\delta\theta)$ rises at small scales and has a turnover at log$\delta\theta$ = $-$2.1 ($\delta\theta$ = 0.5', or a physical scale of 1 kpc). At larger angular scales, the structure function remains rather flat but with a clear bump at log$\delta\theta$=$-$1.4 ($\delta\theta$ = 2.4', or a physical scale of 5.3 kpc). $SF_{\rm RM}(\delta\theta)$ then steepens once again at log$\delta\theta$ = $-$1.1 ($\delta\theta$ = 4.8', or a physical scale of 10 kpc) to a slope of $\sim$ 3.6. 

A striking difference between the L band and the short wavelength RM structure function is the presence of a break at a scale of 1 kpc below which the spectrum steepens to a slope of approximately 1.5. This slope is consistent with 5/3, the value predicted from 3D Kolmogorov turbulence. Since this scale that we are interested in is considerably smaller than the largest angular scale probed, we further divide M51 into 4 quadrants and recompute the structure function in each quadrant. We find that the characteristic break at 1 kpc is present in the RM structure function in each quadrant. Thus, we believe this feature is not merely produced by a localized region, but rather it appears to be present across the entire galaxy. 

We note that the overall amplitude in the flat part of the L band RM structure function is 2$\sigma_{\rm RM}^2$ = 200 rad$^2$ m$^{-4}$, which corresponds to an RM variance of 10 rad m$^{-2}$, is similar to the value computed from RMs of extragalactic sources at high Galactic latitudes in our own Milky Way \citep{mao2010,schnitzeler2010}. This $\sigma_{\rm RM}$ is a factor of 5 smaller than that implied from the short wavelength RM structure function in Section~\ref{section:structure_function_36cm}. This once again reflects the difference in the polarization horizon when observing at different wavelengths at similar resolutions: unlike short wavelength polarized emission that probe through the entire galaxy, L band polarized emission originates from a shallower depth in the galaxy (i.e. the near-side halo) and hence do not probe through the turbulent mid-plane where large fluctuations in electron densities and magnetic fields produce a much larger RM variance.

In order to interpret the RM structure function produced in the less turbulent halo, where random fields no longer dominate over the uniform component, we must first remove any structure function features produced by the large-scale halo magnetic field. In Figure~\ref{fig:sfrm_lband}, we plot the structure function expected from the \cite{fletcher2011} best fit large-scale halo field model as the dashed curve (after subtracting the noise contribution due to uncertainties in the modeled parameters). Although the large-scale halo field alone cannot account for the overall amplitude of the observed RM structure function at L band and it cannot produce the break at $\delta\theta$=0.5', it is evident that the bump in $SF_{\rm RM}(\delta\theta)$ at $\delta\theta$ $\sim$ 2.4' is produced by the large-scale halo magnetic field. We note that the steepening of the structure function at angular lags $>$ 4.8' is not predicted by the Fletcher large-scale halo field. This suggests that there could be an additional large-scale magnetic field component in the halo not modeled by \cite{fletcher2011}. 
 In addition, we plot the structure function expected from turbulence that has an outer scale of log($\delta\theta$)$\sim$$-$2.1 (physical scale of 1 kpc) below which the structure function has a slope of 1.5 and above which the structure function saturates at 2$\sigma_{\rm RM}^2$=10$^2$ rad$^2$ m$^{-4}$ ($\sigma_{\rm RM}$ $\sim$ 7 rad m$^{-2}$) as the dotted line in Figure~\ref{fig:sfrm_lband}. The sum of these two structure functions is denoted as the solid curve in the same Figure. One immediately recognizes that several key features of $SF_{\rm RM}(\delta\theta)$ can be reproduced by the solid curve: (i) the slope of the structure function below $<$ 1 kpc; (ii) the turnover location of the structure function at approximately 1 kpc; (iii) the bump at a scale of $\sim$ 5.3 kpc. We emphasize that this is not a fit to the structure function, but rather a demonstration that the sum of the RM structure function produced by the Fletcher large-scale halo field and that of a three-dimensional Kolmogorov turbulence with an outer scale of 1 kpc can reproduce key features in the observed RM structure function at L band. 
 
We interpret the turnover scale at 1 kpc as the outer scale of turbulence in the magneto-ionic medium in M51's halo.  This outer scale is consistent with the result of our pixel-by-pixel $QU$ fit in Section~\ref{subsection:diffuse_qu_fit}, where we find little wavelength-dependent depolarization across L band. This is because any beam depolarization effects caused by an external inhomogeneous Faraday screen is dominated by the largest fluctuations on scales comparable to the outer scale of turbulence. In our case, the turbulence is spatially resolved (the beam of our L band observations $\sim$ 11" $\ll$ 0.5'), resulting in no/little external Faraday dispersion. 

\subsection{The Source of Turbulence in the Halo of M51}
The implied 1 kpc outer scale of turbulence in the halo of M51 is considerably larger than that inferred in the disk of M51 ($<$ 100 pc). It also exceeds the scales inferred in the Milky Way based on structure functions of extragalactic RMs: few pc in arm regions and $\sim$100 pc in inter-arm regions \citep{haverkorn2006,haverkorn2008}. An increase in the magnetic field correlation length has already been deduced in a number of edge-on galaxies by modeling the decrease in depolarization as a function of height above and below the mid-plane \citep{dumke1995}. Recently, \cite{shneider2014b,shneider2014a} have also derived a larger turbulent correlation length in the halo (215$-$638 pc) than in the disk (40$-$52 pc) in M51 by fitting an analytical depolarization model to observations at 3, 6 and 20 cm. However, the authors acknowledge that the derived halo cell size is likely very uncertain since the fitting algorithm has the assumption that the beam diameter (600pc) largely exceeds the turbulent cell size, which is likely not true in the halo. In galactic disks, HII regions and supernova remnants are thought to be the main drivers of turbulence, while other processes may dominate the energy input in galactic halos \citep[see e.g.,][]{beck1996,elmegreen2004}. In this section, we specifically consider two possible drivers of turbulence in the halo of M51: superbubbles and Parker instability. 

The co-location of a hole in neutral hydrogen and a rotation measure gradient of diffuse polarized emission at L band in the nearby face-on galaxy NGC 6946 \citep{heald2012} hints at a possible connection between energetic events that form superbubbles and structures the magneto-ionic medium. We search for coincidence of RM gradients and HI holes in M51 by using the HI hole catalogue from the THINGS survey \citep{bagetakos2011}. More than 2/3 of the HI holes in the disk of M51 are located in regions devoid of L band polarized emission because holes identified in HI maps are likely in the disk where the density contrast is higher but where synchrotron emission is completely depolarized at L band. We do not find significant RM gradients across HI holes that do fall within the polarized regions. The rarity of such a detection is due to the fact that (1) the HI hole has to be at a certain azimuthal position in order to produce an observable RM gradient, and (2) vertical shearing will destroy polarization signatures of the holes so they have to be in a certain age range. Therefore, the lack of correlation between RM gradient and HI holes should not be taken as an evidence against superbubbles being the energy injection scale in the halo of M51. 
Although the median diameter $\sim$ 700 pc of HI holes in M51 is seemingly too small compared to the outer scale of 1 kpc implied from our structure function analysis, this can be reconciled by taking into account that hot gas parcels from supernova remnants and superbubbles would expand as they rise above/below the mid-plane. In the recent numerical simulation of the magnetized ISM of \cite{gent2013}, the authors found an increase in the correlation length scale by a factor of 1.5$-$3 between the mid-plane and 800 pc above the plane. This suggests that by the time the bubbles rise to the top of the synchrotron disk at L band, their characteristic size would be at least 1 kpc, which is consistent with the outer scale found in our structure function analysis. Therefore, the observed 1 kpc magnetized halo correlation length scale is consistent with superbubbles driving turbulence in the halo of M51. However, to definitively draw this conclusion, we require wide-band polarimetry at S band (2-4 GHz) to probe closer to the turbulent mid-plane (z $<$ 1.2 kpc). If one finds a smaller turbulent correlation length of the magnetized medium and more coincidence between HI holes and RM gradients at S band, then it will strengthen the argument for superbubbles being the main energy driver in the halo and producing imprint onto the RM structure function. 

Another plausible source of turbulence in the halo of M51 is the Parker instability\footnotemark[13]\footnotetext[13]{also known as magnetic buoyancy instability.} \citep{parker1966} which is the instability experienced by a purely horizontal magnetic field under vertical perturbations. Cosmic ray pressure inflates magnetic field lines into the halo causing material to slide down the field lines, which in turn further inflates the field due to buoyancy. In fact, \cite{parker1992} suggested that the entire surface of a galactic disk (on both sides) is packed with Parker loops with a characteristic scale of 0.1$-$1 kpc. Three-dimensional MHD simulations by \cite{kim2002} found the dominant growing mode to be 10-20 times the scale height of the gas disk in the direction of the magnetic field. For M51, which has an HI scale height of 160 pc \citep{bagetakos2011}, Parker loops in the galaxy can have a characteristic scale of $\sim$ 1-3 kpc. If these magnetized loops thread the warm ionized medium of M51, then they can produce RM sign reversals (after the Galactic foreground and the large-scale magnetic fields have been accurately subtracted from the observed RM) and leave imprints onto the long wavelength (L band) RM structure function\footnotemark[14]\footnotetext[14]{We suggest that it is much easier to search for signatures of Parker loops in RM maps produced by long wavelength data because only the near-side halo of the galaxy is visible in polarization. At shorter wavelengths, polarized emission probes through the turbulent mid-plane, and thus relatively weak features produced by Parker instabilities would be completely washed out by strong RM fluctuations in the mid-plane. In fact, \cite{fletcher2011} were not able to find any periodic RM pattern in the short wavelength (3 and 6 cm) RM map and took this as a lack of evidence for Parker loops.}. We note that it is challenging to identify by eye positive and negative RM patches separated by $\sim$ 1 kpc in our L band RM map because there exists fluctuations on smaller scales that are superposed on top of this positive/negative pattern. 

Since the inferred outer scale in M51's halo matches well with the expected size of Parker loops, we propose that the 1 kpc outer scale inferred from the L band RM structure function could be produced by Parker instabilities. Turbulence in the halo can be driven by Parker instability and then cascade down to smaller scales via inherit short wavelength structures perpendicular to the initial field direction as well as induced shear flow instabilities \citep[e.g.,][]{kim2001}. In theory, these short wavelength structures should be visible in both the Faraday depth map and the structure function. However, the wavelength of this mode is often very small  (much smaller than the disk scale height, in the case of M51 $<<$ 160pc) \citep{kim2002} and hence it cannot be resolved by our L band observations which have a resolution of  370 pc. If Parker instability indeed exists, then high resolution polarization observations should reveal very small scale RM fluctuations perpendicular to the direction of the large-scale fields. While a magnetic field reversal along an arc feature in the halo of M31 had been interpreted as a Parker instability loop anchored in a massive HI cloud \citep{beck1989}, the evidence that we present in this work is the first to suggest that Parker loops could be present across the entire galactic halo, and not just in a single loop associated with an individual gas cloud. We note that our argument presented here is solely based on the characteristic scale of RM fluctuation. As future work, comparing mock Faraday rotation maps and structure functions generated from numerical simulations of Parker instability with wide-band polarization observations is crucial to further constrain properties of turbulence in galactic halos.

If the observed 1 kpc correlation length in the halo of M51 indeed corresponds to Parker instability loops, it has important implications on the large-scale magnetic field amplification mechanism in galaxies.
Inflated magnetized loops in the halo together with magnetic reconnections can lead to the fast cosmic-ray-driven dynamo effect \citep{parker1992,hanasz2004}. This more efficient $\alpha$-effect can lead to much shorter growth time than the classical $\alpha$-$\omega$ dynamo, on times scales of few hundred Myrs. This process will allow coherent magnetic fields to be established in young galaxies, as well as galaxies undergoing tidal interactions, such as the Magellanic Clouds \citep{gaensler2005,mao2008,mao2012b} and M51 itself. Parker loops allow the transportation of magnetic flux into the halo and may be the key ingredient for the existence of large-scale magnetic fields in galactic halos.

\section{Conclusions and Future Work}
\label{section:summary}
In this paper, we have presented new wide-band, multi-configuration Jansky VLA polarization observations of M51 at L band which have the best $\lambda^2$ coverage in this frequency range to date. For the first time, we have fitted the observed diffuse polarized emission from an external galaxy as a function of $\lambda^2$ directly to various models of the Faraday rotating and synchrotron emitting medium. The majority of sight lines through M51 do not exhibit $\lambda$-dependent depolarization across L band and therefore their $Q(\lambda^2)$ and $U(\lambda^2)$ are well fitted by Faraday rotation occurring in an external uniform screen. This is consistent with the picture of polarized emission at L band being originated from the top layer of the synchrotron emitting disk and then Faraday rotated in the thermal halo. Using the deduced relative extent of the synchrotron emitting gas and the thermal gas (warm ionized medium), the synchrotron cooling time scale, and a cosmic ray diffusion coefficient similar to that in the Milky Way, the warm ionized medium scale height in M51 is estimated to be at least 1.2 kpc. The predicted Faraday depth distribution from the \cite{fletcher2011} bisymmetric halo magnetic field model at L band and our observed RMs can be reconciled by introducing an additional vertical coherent magnetic field component (RM = $-$9 rad m$^{-2}$) in the halo, which could be of primordial or dynamo origin. With the improved wavelength coverage of our observations and hence a narrower rotation measure spread function with lower side-lobe levels, we do not detect any polarized emission from the far-side halo at Faraday depths of $\sim$ $\pm$ 200 rad m$^{-2}$ as suggested by \cite{braun2010}. This non-detection implies that the turbulent mid-plane must have an RM variance that exceeds 20 rad m$^{-2}$, which is consistent with the short wavelength (3 cm and 6 cm) RM variance. The RM structure function derived using 3 and 6 cm data (which probe through the entire galaxy) shows no break on small angular scales, implying that the energy injection scale in M51's disk is smaller than the resolution of the observation ($<$ 560 pc). On the other hand, the RM structure function of L band data (which only probe through the halo) has a break at 1 kpc. Both superbubble-driven and Parker instability-driven turbulence could produce correlation lengths of approximately 1 kpc in the halo of M51. 

Throughout this paper, we have demonstrated the importance of wide-band polarization data when extracting properties of magnetic fields and turbulence in the underlying magneto-ionic medium. Future wide-band polarization observations between 6 cm and 18 cm (S band) can probe deeper into the galaxy and may reveal connections between large-scale fields in the disk and in the halo and possibly reveal a stronger correlation between HI holes and the RM distribution. We can confirm or rule out the existences of Parker loops in M51's halo by obtaining even higher spatial resolution L band polarization data to search for short wavelength fluctuations in Faraday rotation perpendicular to the direction of the large-scale magnetic field. Finally, to understand global large-scale halo magnetic field configurations and to evaluate the universality of superbubble and Parker instability-driven turbulence in galactic halos, a systematic polarization survey of mildly inclined galaxies at L and S bands would be of crucial importance. 

\appendix 

\section{Models for the Faraday Rotating Medium used for the Direct $Q$ $U$ Fits}
\label{appendix:A}

We summarize in the different models which are used to fit the observed $Q$/$I$ and $U$/$I$ of extragalactic sources and the diffuse polarized emission. 

First, we consider the simplest model where intrinsic polarized background emission with a fractional polarization ${p_0}$ and an intrinsic polarization angle ${\phi_0}$ propagates through a uniform foreground screen which has a rotation measure RM. In this case, the observed fractional complex polarization $\vec{p}$ is
\begin{equation}
\vec{p} = {p_0}e^{2i(\phi_0+\rm{RM}\lambda^2)}. 
\end{equation}

In the case of an inhomogeneous foreground screen, under the assumption that the turbulent cell size is much smaller than the beam size, polarized intensity is reduced due to the varying RM within the beam by the so-called external Faraday dispersion effect \citep{burn1966,tribble1991}. The fractional complex polarization becomes
\begin{equation}
\vec{p} = {p_0}e^{-2\sigma_{\rm RM}^2\lambda^4}e^{2i(\phi_0+\rm{RM}\lambda^2)}, 
\end{equation}
where the RM variance $\sigma_{\rm RM}$ quantifies the fluctuations in electron densities and magnetic fields within the beam.

If only part of the background polarized radiation passes through the foreground inhomogeneous screen, the complex polarization depends on the fraction of the background source covered by the inhomogeneous screen $f_c$, 
\begin{equation}
\vec{p} = {p_0}[f_c e^{-2\sigma_{\rm RM}^2\lambda^4}+(1-f_c)]e^{2i(\phi_0+\rm{RM}\lambda^2)}. 
\end{equation}

We then consider the case where there is differential depolarization intrinsic to the background source (the classical Burn slab with a total Faraday depth R, \cite{burn1966}) and the radiation then pass through a uniform foreground screen with a Faraday rotation of RM 
\begin{equation}
\vec{p} = {p_0}\frac{{\rm sin}R\lambda^2}{R\lambda^2}e^{2i(\phi_0+\frac{1}{2}R\lambda^2+\rm{RM}\lambda^2)}. 
\end{equation}

If there is internal Faraday dispersion occurring within the background source characterized by $\sigma_{\rm RM}$, then the expected complex fractional polarization is:
\begin{equation}
\vec{p} = {p_0}\frac{1-e^{-S}}{S}e^{2i(\phi_0+\rm{RM}\lambda^2)},~\rm where~S=2\lambda^4\sigma_{RM}^2-2i\lambda^2R
\end{equation}

Finally, there is the possibility that the observed polarized emission is the sum of two different polarization components, A and B: 
\begin{equation}
\vec{p} = {p_{0,A}}e^{2i(\phi_{0,A}+\rm{RM_A}\lambda^2)}+{p_{0,B}}e^{2i(\phi_{0,B}+\rm{RM_B}\lambda^2)}. 
\end{equation}

 {\it Facilities:} The Very Large Array
The National Radio Astronomy Observatory is a facility of the National Science Foundation operated under cooperative agreement by Associated Universities, Inc.

The authors thank Rainer Beck and Henrik Junklewitz for a close reading of the manuscript. The authors acknowledge fruitful discussions with Judith Irwin, Bob Lindner, Snezana Stanimirovi\'{c}, Shane O'Sullivan, Jay Gallagher, Tony Wong, Cathy Horellou, Chris Hales, Amanda Kepley, Jamie Farnes, Blakesley Burkhart, Xiao Hui Sun, Larry Rudnick, Peter Tribble, Steve Spangler, George Heald, Anvar Shukurov, Tim Robishaw, Rick Perley, Aritra Basu and Walter Max-Moerbeck.

\clearpage
\begin{deluxetable*}{lllc}
\tablecolumns{4} 
\tablewidth{0pc} 
\tablecaption{Coordinates of polarized extragalactic background sources in the field of view} 
\vspace{0.01in}
\tablehead{   
\colhead{Source Name} &     \colhead{RA(J2000)(hms)}   &    \colhead{DEC(J2000)(dms)} &   \colhead{Distance from the center of FOV(')} 
}
\startdata 
J1330+4703a & 13:30:45.085 &  $+$47:03:08.76  & 12.1 \\
J1330+4703b & 13:30:45.286 &  $+$47:03:28.76  & 11.9 \\
J1329+4658a* & 13:29:32.315 &  $+$46:58:45.38  & 13.2 \\
J1329+4658b* & 13:29:28.796 &  $+$46:58:49.34  & 13.3\\
J1329+4658c* & 13:29:39.153 &  $+$46:59:09.45  & 12.5\\
J1331+4713a* & 13:31:22.554 &  $+$47:13:21.32  & 15.3 \\
J1331+4713b* & 13:31:27.452 &  $+$47:13:01.08  & 16.1\\
J1330+4710   & 13:30:15.978 &  $+$47:10:23.36  &  4.0 \\
J1329+4717*   & 13:29:41.628 &  $+$47:17:35.46  &  6.5\\
J1330+4730   & 13:30:32.513 &  $+$47:30:55.08  & 20.6\\
J1329+4706   & 13:29:31.089  &  $+$47:6:27.37 & 6.2 \\

 \enddata
 \label{table:egs_catalog}
\end{deluxetable*}

\clearpage
\LongTables
\begin{landscape}
\begin{deluxetable}{llllllllllllll}
\label{table:pointsource_RM_fit}
\tablecolumns{14} 
\tabletypesize{\tiny}
\tablewidth{0pc} 
\tablecaption{Results of Stokes $QU$ fit for all polarized extragalactic sources within 20' of the field center.} 
\vspace{-0.8in}
\tablehead{   
\colhead{Source Name} &  \colhead{Model}   &   \colhead{p$_{0,A}$} &   \colhead{$\phi_{0,A}$} &  \colhead{RM$_A$} &  \colhead{$\sigma_{\rm RM}$}  & \colhead{f$_c$} & \colhead{R}   & \colhead{p$_{0,B}$} &   \colhead{$\phi_{0,B}$} &  \colhead{RM$_B$}  & \colhead{$\chi^2_r$} &\colhead{DOF} & \colhead{BIC}
\\
\colhead{} &  \colhead{}   &   \colhead{(fraction)} &   \colhead{(rad)} &  \colhead{(rad m$^{-2}$)} &  \colhead{(rad m$^{-2}$)} &  \colhead{(fraction)} & \colhead{(rad m$^{-2}$)}   & \colhead{(fraction)} &   \colhead{(rad)} &  \colhead{(rad m$^{-2}$)}  & \colhead{} &\colhead{} & \colhead{}
}
\startdata 
J1330$+$4703a & i &      0.067 $\pm$ 0.002 &       4.60 $\pm$ 0.05 &       $+$17. $\pm$ 1. & - & - &- &- &- &- &      2.727 &        87. &      -524. \\
& {\textbf {ii}} &      0.095 $\pm$ 0.004 &       4.57 $\pm$ 0.04 &      +17.2 $\pm$ 0.8 &        8.4 $\pm$ 0.6 & - &- &- &- & - &      1.584 &        86. &      -621. \\
 & iii &       0.11 $\pm$ 0.01 &       4.58 $\pm$ 0.04 &      +17.0 $\pm$ 0.8 &        16. $\pm$ 3. &       0.58 $\pm$ 0.05 &- &- &- & - &      1.527 &        85. &      -624. \\
 & iv &      0.092 $\pm$ 0.004 &       4.57 $\pm$ 0.04 &        +4. $\pm$ 1. &- &- &       +27. $\pm$ 2. &- &- &- &      1.611 &        86. &      -619. \\
 & v &      0.098 $\pm$ 0.005 &       4.50 $\pm$ 0.05 &       +12. $\pm$ 4. &        11. $\pm$ 3. &- &       +14. $\pm$ 9. &- &- &- &      1.544 &        85. &      -622. \\
 & vi &      0.067 $\pm$ 0.002 &       4.59 $\pm$ 0.05 &       +17. $\pm$ 1. &- & - &-  &      0.007 $\pm$ 0.002 &        3.7 $\pm$ 0.3 &      +153. $\pm$ 7. &  2.344 &        84. &      -553. \\
J1330+4703b & {\bf i} &      0.046 $\pm$ 0.003 &        0.2 $\pm$ 0.1 &       +10. $\pm$ 2. & - & - &- &- &- &- &      2.235 &        87. &      -459. \\
& ii &      0.046 $\pm$ 0.003 &        0.2 $\pm$ 0.1 &       +10. $\pm$ 2. &         0. $\pm$ 4. & - &- &- &- & - &      2.261 &        86. &      -455. \\
& iii &      0.046 $\pm$ 0.003 &        0.2 $\pm$ 0.1 &       +10. $\pm$ 2. &         0. $\pm$ 0. &         1. $\pm$ 0. &- &- &- & - &      2.287 &        85. &      -455. \\
& iv &      0.046 $\pm$ 0.003 &        0.2 $\pm$ 0.1 &       +10. $\pm$ 7. &- &- &        -0. $\pm$ 10 &- &- &- &      2.261 &        86. &      -455. \\
 & v &      0.046 $\pm$ 0.003 &        0.2 $\pm$ 0.1 &       +10. $\pm$ 7. &         0. $\pm$ 5. &- &        +0. $\pm$ 10 &- &- &- &      2.287 &        85. &      -452. \\
 & vi &      0.005 $\pm$ 0.003 &        -5. $\pm$ 1. &      +112. $\pm$ 20 &- & - &-  &      0.045 $\pm$ 0.003 &        0.2 $\pm$ 0.2 &       +11. $\pm$ 3. &      2.222 &        84. &      -456. \\
J1329+4658a & i &       0.24 $\pm$ 0.01 &        0.3 $\pm$ 0.1 &       +13. $\pm$ 2. & - & - &- &- &- &- &      3.171 &        87. &      -117. \\
&{\textbf {ii}} &       0.47 $\pm$ 0.06 &        0.2 $\pm$ 0.1 &       +16. $\pm$ 2. &        12. $\pm$ 1. & - &- &- &- & - &      2.375 &        86. &      -185. \\
 & iii &         1. $\pm$ 0. &       0.26 $\pm$ 0.08 &       +15. $\pm$ 2. &        26. $\pm$ 2. &       0.82 $\pm$ 0.02 &- &- &- & - &      2.178 &        85. &      -202. \\
 & iv &       0.41 $\pm$ 0.04 &        0.2 $\pm$ 0.1 &        -1. $\pm$ 2. &- &- &       +34. $\pm$ 3. &- &- &- &      2.449 &        86. &      -179. \\
 & v &        0.6 $\pm$ 0.1 &        0.0 $\pm$ 0.2 &       +13. $\pm$ 4. &        20. $\pm$ 5. &- &       +22. $\pm$ 10 &- &- &- &      2.281 &        85. &      -192. \\
 & vi &         1. $\pm$ 0. &       -0.6 $\pm$ 0.1 &       +27. $\pm$ 3. &- & - &-  &       0.83 $\pm$ 0.02 &        0.6 $\pm$ 0.2 &       +32. $\pm$ 4. &      2.280 &        84. &      -190. \\
J1329+4658b & i &       0.16 $\pm$ 0.01 &        5.2 $\pm$ 0.2 &       +11. $\pm$ 3. & - & - &- &- &- &- &      2.889 &        87. &      -139. \\
 &{\textbf {ii} }&       0.45 $\pm$ 0.07 &        5.3 $\pm$ 0.1 &        +9. $\pm$ 3. &        14. $\pm$ 1. & - &- &- &- & - &      2.005 &        86. &      -214.\\
& iii &        0.7 $\pm$ 0.2 &        5.3 $\pm$ 0.1 &        +9. $\pm$ 2. &        21. $\pm$ 3. &       0.86 $\pm$ 0.03 &- &- &- & - &      1.926 &        85. &      -219. \\
 & iv &       0.35 $\pm$ 0.04 &        5.2 $\pm$ 0.1 &        -9. $\pm$ 3. &- &- &       +38. $\pm$ 2. &- &- &- &      2.104 &        86. &      -206. \\
 & v &        0.7 $\pm$ 0.2 &        5.8 $\pm$ 0.2 &        +8. $\pm$ 4. &        23. $\pm$ 6. &- &       -42. $\pm$ 10 &- &- &- &      1.882 &        85. &      -223. \\
 & vi &        0.5 $\pm$ 0.5 &        5.0 $\pm$ 0.3 &        -7. $\pm$ 7. &- & - &-  &        0.6 $\pm$ 0.5 &        6.1 $\pm$ 0.3 &        -0. $\pm$ 6. &      2.060 &        84. &      -206. \\
J1329+4658c & i &      0.124 $\pm$ 0.003 &       0.77 $\pm$ 0.05 &       +16. $\pm$ 1. & - & - &- &- &- &- &     35.119 &        87. &      2163. \\
& ii &      0.202 $\pm$ 0.005 &       0.76 $\pm$ 0.02 &      +16.7 $\pm$ 0.4 &       10.3 $\pm$ 0.2 & - &- &- &- & - &      5.118 &        86. &      -449. \\
 & {\textbf  {iii} }&      0.258 $\pm$ 0.009 &       0.76 $\pm$ 0.01 &      +16.6 $\pm$ 0.3 &       18.3 $\pm$ 0.7 &       0.69 $\pm$ 0.01 &- &- &- & - &      2.621 &        85. &      -662. \\
 & iv &      0.189 $\pm$ 0.004 &       0.76 $\pm$ 0.02 &       +1.1 $\pm$ 0.6 &- &- &      +31.3 $\pm$ 0.7 &- &- &- &      6.408 &        86. &      -338. \\
& v &      0.218 $\pm$ 0.005 &       0.68 $\pm$ 0.03 &       +14. $\pm$ 1. &       16.0 $\pm$ 0.7 &- &       +11. $\pm$ 4. &- &- &- &      3.747 &        85. &      -567. \\
& vi &      0.121 $\pm$ 0.003 &       0.78 $\pm$ 0.05 &       +16. $\pm$ 1. &- & - &-  &      0.011 $\pm$ 0.003 &        0.3 $\pm$ 0.4 &      -181. $\pm$ 8. &     30.130 &        84. &      1650. \\
J1331+4713a & i &      0.106 $\pm$ 0.001 &       0.58 $\pm$ 0.03 &      +12.1 $\pm$ 0.5 & - & - &- &- &- &- &      6.655 &        87. &      -306. \\
 & {\textbf {ii} }&      0.131 $\pm$ 0.003 &       0.64 $\pm$ 0.02 &      +10.7 $\pm$ 0.4 &        6.9 $\pm$ 0.3 & - &- &- &- & - &      2.708 &        86. &      -648. \\
 & iii &      0.131 $\pm$ 0.005 &       0.64 $\pm$ 0.02 &      +10.7 $\pm$ 0.4 &         8. $\pm$ 4. &        0.9 $\pm$ 0.7 &- &- &- & - &      2.739 &        85. &      -645. \\
 & iv &      0.129 $\pm$ 0.002 &       0.64 $\pm$ 0.02 &       -0.7 $\pm$ 0.7 &- &- &      +22.8 $\pm$ 0.9 &- &- &- &      2.716 &        86. &      -648. \\
& v &      0.132 $\pm$ 0.003 &       0.69 $\pm$ 0.02 &       +16. $\pm$ 3. &         8. $\pm$ 2. &- &       -14. $\pm$ 6. &- &- &- &      2.623 &        85. &      -655. \\
& vi &       0.04 $\pm$ 0.01 &        0.5 $\pm$ 0.3 &        -2. $\pm$ 5. &- & - &-  &       0.11 $\pm$ 0.01 &       0.83 $\pm$ 0.09 &       +11. $\pm$ 2. &      2.587 &        84. &      -656. \\
J1331+4713b & i &      0.117 $\pm$ 0.002 &       0.28 $\pm$ 0.02 &       +6.4 $\pm$ 0.5 & - & - &- &- &- &- &      5.985 &        87. &      -316. \\
& {\textbf {ii} }&      0.127 $\pm$ 0.005 &       0.30 $\pm$ 0.03 &       +6.0 $\pm$ 0.5 &         4. $\pm$ 1. & - &- &- &- & - &      5.725 &        86. &      -340. \\
 & iii &      0.127 $\pm$ 0.005 &       0.30 $\pm$ 0.03 &       +6.0 $\pm$ 0.5 &         4. $\pm$ 1. &         1. $\pm$ 0. &- &- &- & - &      5.792 &        85.  &      -338. \\
 & iv &      0.127 $\pm$ 0.005 &       0.30 $\pm$ 0.03 &       +13. $\pm$ 2. &- &- &       -15. $\pm$ 3. &- &- &- &      5.718 &        86. &      -341. \\
& v &      0.129 $\pm$ 0.006 &       0.32 $\pm$ 0.03 &       +11. $\pm$ 4. &        -5. $\pm$ 3. &- &       -10. $\pm$ 7. &- &- &- &      5.727 &        85. &      -342. \\
& vi &      0.011 $\pm$ 0.007 &        1.5 $\pm$ 0.6 &       -29. $\pm$ 20 &- & - &-  &      0.111 $\pm$ 0.007 &       0.28 $\pm$ 0.09 &        +7. $\pm$ 2. &      5.347 &        84. &      -376. \\
J1330+4710 & {\bf i } &     0.0762 $\pm$ 0.0007 &       4.01 $\pm$ 0.02 &      +26.0 $\pm$ 0.4 & - & - &- &- &- &- &      1.113 &        87. &      -750. \\
 & ii &      0.082 $\pm$ 0.002 &       4.01 $\pm$ 0.02 &      +26.1 $\pm$ 0.4 &        3.7 $\pm$ 0.5 & - &- &- &- & - &      0.962 &        86. &      -760. \\
& iii &      0.082 $\pm$ 0.002 &       4.01 $\pm$ 0.02 &      +26.1 $\pm$ 0.4 &        3.7 $\pm$ 0.5 &         1. $\pm$ 0. &- &- &- & - &      0.973 &        85. &      -756. \\
& iv &      0.082 $\pm$ 0.002 &       4.01 $\pm$ 0.02 &      +19.7 $\pm$ 0.9 &- &- &       +13. $\pm$ 2. &- &- &- &      0.961 &        86. &      -760. \\
 & v &      0.082 $\pm$ 0.002 &       4.00 $\pm$ 0.02 &       +22. $\pm$ 3. &         4. $\pm$ 2. &- &        +8. $\pm$ 6. &- &- &- &      0.961 &        85. &      -757. \\
 & vi &      0.074 $\pm$ 0.002 &       4.05 $\pm$ 0.03 &      +25.0 $\pm$ 0.7 &- & - &-  &      0.005 $\pm$ 0.002 &        2.6 $\pm$ 0.4 &       +64. $\pm$ 9. &      0.851 &        84. &      -764. \\
J1329+4717 & i &      0.111 $\pm$ 0.004 &       5.71 $\pm$ 0.07 &       +23. $\pm$ 1. & - & - &- &- &- &- &      2.555 &        87. &      -394. \\
& {\textbf {ii}} &       0.19 $\pm$ 0.01 &       5.70 $\pm$ 0.04 &      +23.5 $\pm$ 0.9 &       10.3 $\pm$ 0.5 & - &- &- &- & - &      1.127 &        86. &      -515. \\
& iii &       0.25 $\pm$ 0.03 &       5.70 $\pm$ 0.04 &      +23.5 $\pm$ 0.8 &        19. $\pm$ 2. &       0.70 $\pm$ 0.03 &- &- &- & - &      1.013 &        85.  &      -522. \\
 & iv &      0.176 $\pm$ 0.008 &       5.70 $\pm$ 0.04 &        +8. $\pm$ 1. &- &- &       +31. $\pm$ 1. &- &- &- &      1.185 &        86. &      -510. \\
 & v &       0.21 $\pm$ 0.01 &        5.7 $\pm$ 0.1 &       +24. $\pm$ 3. &        17. $\pm$ 1. &- &        -2. $\pm$ 10 &- &- &- &      1.085 &        85. &      -516. \\
 & vi &        0.8 $\pm$ 0.9 &       4.97 $\pm$ 0.09 &       +19. $\pm$ 6. &- & - &-  &        0.8 $\pm$ 0.9 &       6.41 $\pm$ 0.09 &       +21. $\pm$ 6. &      1.136 &        84. &      -509. \\
J1329+4706 & i &      0.172 $\pm$ 0.009 &        29. $\pm$ 6. &       +22. $\pm$ 2. & - & - &- &- &- &- &      1.703 &  87. &      -265. \\
 & {\textbf {ii}} &       0.36 $\pm$ 0.03 &        32. $\pm$ 4. &       +21. $\pm$ 2. &       12.0 $\pm$ 0.9  & - &- &- &- & - &      1.038 &        86. &      -320. \\
 & iii &       0.51 $\pm$ 0.09 &       3.69 $\pm$ 0.07 &       +21. $\pm$ 2. &        20. $\pm$ 3. &       0.79  $\pm$ 0.03 &- &- &- & - &      0.980 &        85. &      -322. \\
 & iv &       0.31 $\pm$ 0.02 &        32. $\pm$ 5. &        +3. $\pm$ 2. &- &- &       +35. $\pm$ 2. &- &- &- &      1.082 &        86. &      -316. \\
 & v &       0.43 $\pm$ 0.06 &        42. $\pm$ 10 &       +22. $\pm$ 3. &        20. $\pm$ 3. &- &       -16.  $\pm$ 10 &- &- &- &      0.998 &        85. &      -321. \\
 & vi &         3. $\pm$ 9. &       176. $\pm$ 10 &       +11. $\pm$ 6. &- & - &-  &         3. $\pm$ 9. &       4.56 $\pm$ 0.07 &       +12. $\pm$ 3. &      1.047 &        84. &      -314. \\
\enddata
\end{deluxetable} 
\clearpage
\end{landscape}


\begin{deluxetable*}{lllll}
\tablecolumns{4} 
\tablewidth{0pc} 
\tablecaption{Properties of extragalactic polarized background sources that are in both this work and \cite{farnes2013}} 
\vspace{0.01in}
\tablehead{   
\colhead{Source Name} &     \colhead{FD}   &    \colhead{Fractional Polarization} &   \colhead{FD} &   \colhead{Fractional Polarization}  \\
\colhead{} &     \colhead{at L band}   &    \colhead{at L band} &   \colhead{at 610 MHz} &   \colhead{at 610 MHz} \\
\colhead{} &     \colhead{(rad m$^{-2}$)}   &    \colhead{} &   \colhead{(rad m$^{-2}$)} &   \colhead{} 
}
\startdata 
J1330+4703a &  $+$16.6 $\pm$ 0.3 & 0.14 $\pm$ 0.01 & $+$11.15 $\pm$ 0.05 & 0.0172   $\pm$ 0.0009 \\
J1330+4703b &  $+$26.0 $\pm$ 0.4 &  0.0762 $\pm$  0.0007 &  $+$33.52 $\pm$ 0.03 & 0.018   $\pm$ 0.002 \\
J1329+4658c &  $+$17.2 $\pm$ 0.8 & 0.076 $\pm$  0.004 & $-$3.240 $\pm$ 0.04 &  0.020 $\pm$  0.002 \\
J1331+4713a &  $+$10 $\pm$ 2 & 0.046 $\pm$ 0.003 & $-$15.97 $\pm$  0.03 &  0.027  $\pm$ 0.002 \\
J1331+4713b &  $+$10.7 $\pm$ 0.4 & 0.112 $\pm$  0.003 & $+$11.51 $\pm$ 0.03  & 0.068  $\pm$ 0.002 \\
J1330+4710   &  $+$6.0 $\pm$ 0.5 & 0.119  $\pm$ 0.006 &  $+$7.67 $\pm$ 0.04 & 0.072  $\pm$      0.003
 \enddata
 \label{table:overlapped_sample}
\end{deluxetable*} 

\begin{figure*}
\centering
\includegraphics[width=1.0\textwidth]{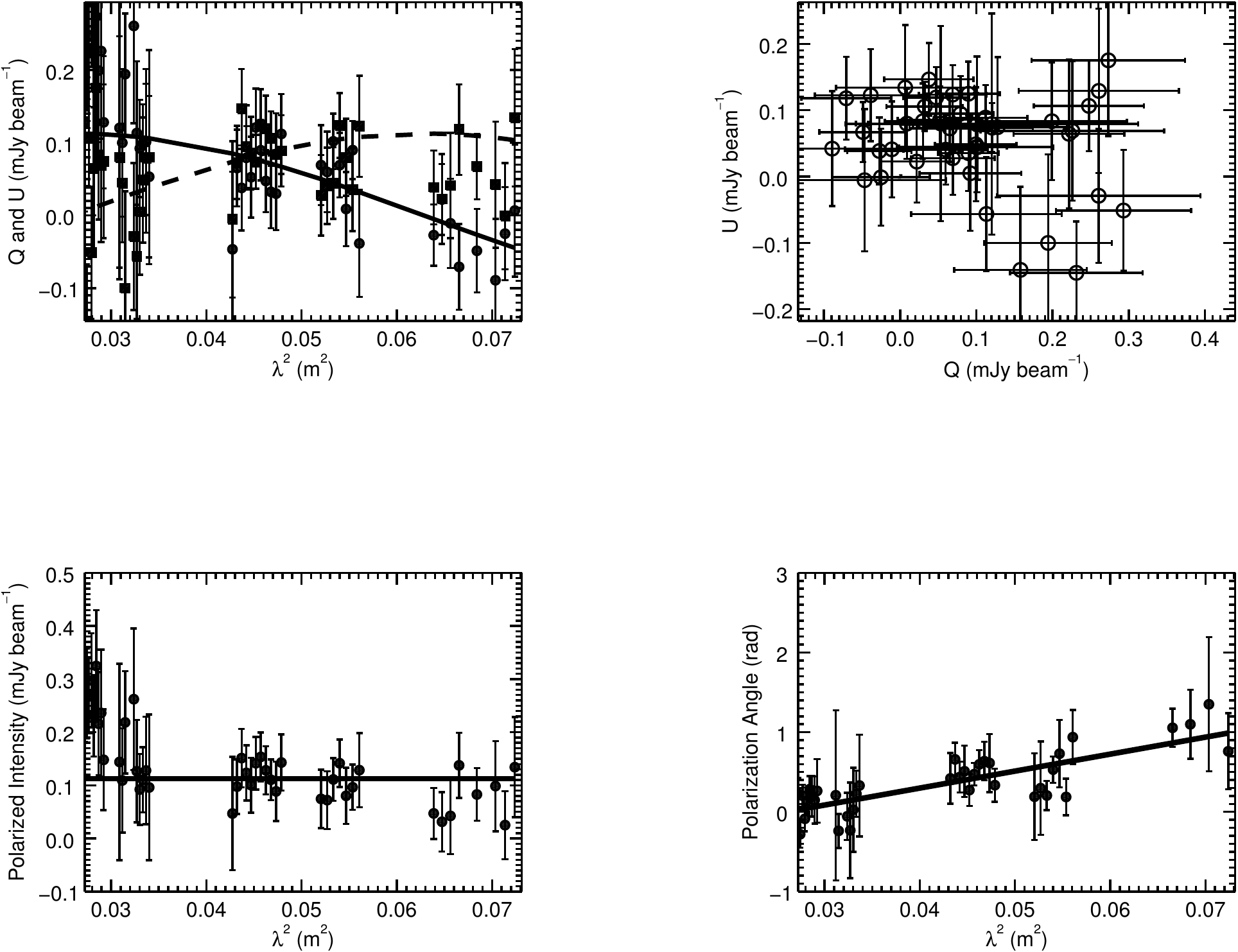}
\caption{Polarized emission along a sight line towards M51 and its corresponding best fit (solid line) to the uniform external Faraday screen model. This figure has the same layout as in Figure~\ref{fig:pt_source_1}.}
\label{fig:}
\end{figure*}

\begin{figure*}
\centering
\includegraphics[width=1.0\textwidth]{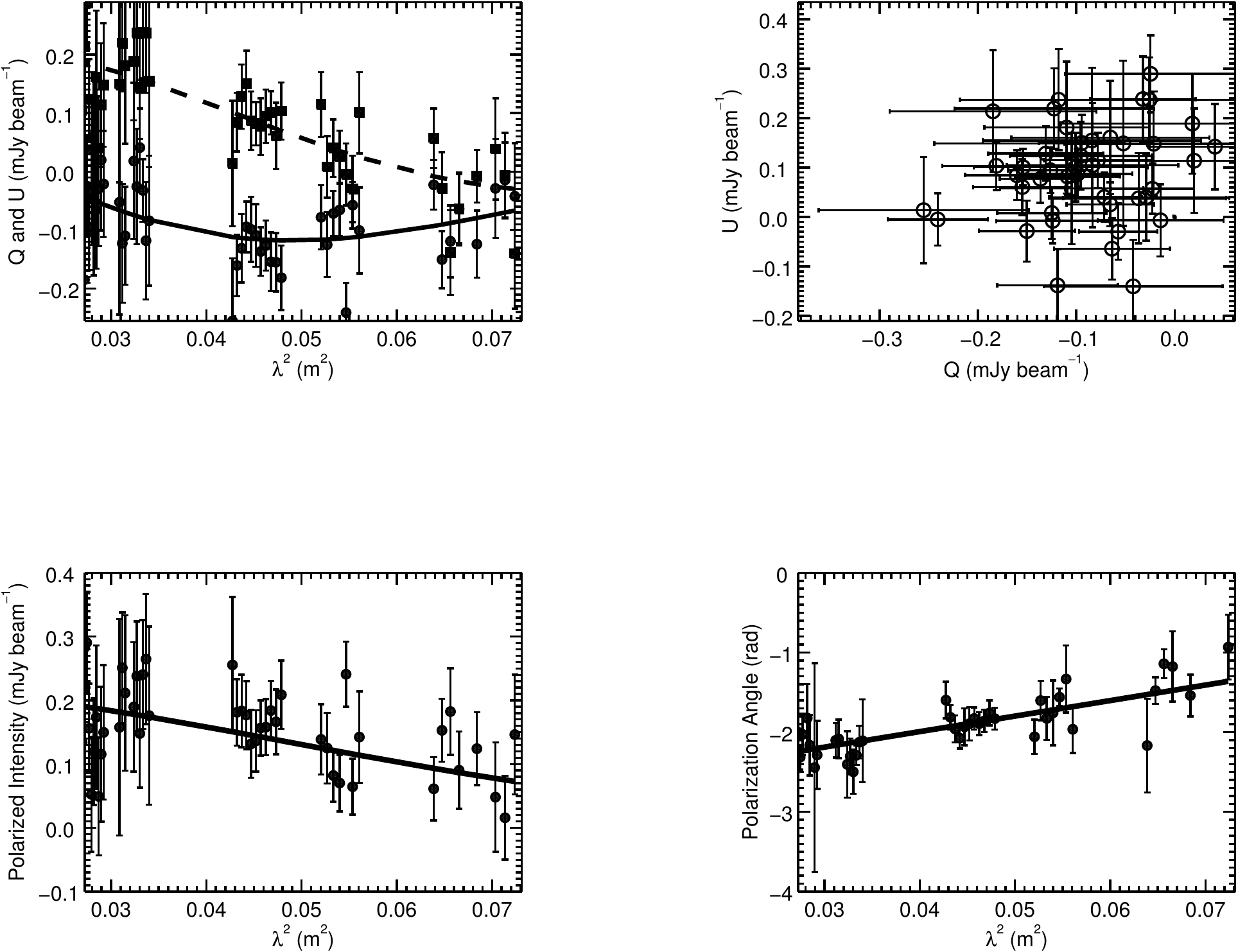}
\caption{Polarized emission along a sight line towards M51 and its corresponding best fit (solid line) to the inhomogeneous external Faraday screen model. This figure has the same layout as in Figure~\ref{fig:pt_source_1}.}
\label{fig:}
\end{figure*}

\clearpage

\end{document}